\documentclass{emulateapj}
\usepackage{rotating}

\shorttitle{Brown Dwarf Disks at Ages of 5-10 Myr}
\shortauthors{Riaz, Lodieu \& Gizis}

\begin{document}

\title{Brown Dwarf Disks at Ages of 5-10 Myr}

\author{Basmah Riaz\altaffilmark{1,2}, Nicolas Lodieu\altaffilmark{1,2} \& John E. Gizis\altaffilmark{3}}
\altaffiltext{1}{Instituto de Astrof'sica de Canarias, E-38200 La Laguna, Tenerife, Spain}
\altaffiltext{2}{Departamento de Astrof\'isica, Universidad de La Laguna, E-38205 La Laguna, Tenerife, Spain}
\altaffiltext{3}{Department of Physics and Astronomy, University of Delaware, Newark, DE 19716}

\email{basmah@iac.es}

\begin{abstract}
We present {\it Spitzer}/IRAC and MIPS 24 $\micron$ archival observations for 28 brown dwarfs in the Upper Scorpius (UppSco) region. We find a disk fraction of 10.7\%$^{+8.7\%}_{-3.3\%}$. One object shows a small excess at 24 $\micron$ but none at shorter wavelengths, and may be a candidate transition disk. Three objects show emission in the 10$\micron$ silicate feature and we present compositional fits for these sources. Flat structures are observed for all disk sources in UppSco. Also presented are the MIPS/70$\micron$ observations for the TW Hydrae Association brown dwarf 2MASS J1139511-315921. We discuss the structure and chemistry of brown dwarf disks at ages of $\sim$5-10 Myr, and consider the possible effects of the brown dwarf densities in these clusters on the disk lifetimes. 

\end{abstract}

\keywords{circumstellar matter -- stars: low-mass, brown dwarfs -- open clusters and associations: individuals (Upper Scorpius, TW Hydrae Association) -- stars: individual (2MASS J1139511-315921)}

\section{Introduction}
The Upper Scorpius (UppSco) OB association is part of the Scorpius Centaurus complex (de Geus et al.\ 1989), located at a distance of 145$\pm$2 pc from the Sun (de Bruijne et al.\ 1997). The age of USco is $\sim$5 Myr, with little scatter of $\pm$3 Myr (e.g., Preibisch \& Zinnecker 1999; Slesnick et al. 2008). The region is relatively free of extinction (Av $\leq$ 2 mag) and star formation has already ended (Walter et al.\ 1994). The association has been targeted at multiple wavelengths over the past decade, starting with the X-rays (e.g.\ Preibisch et al.\ 1998) and with Hipparcos (de Zeeuw et al.\ 1999), as well as in the optical (Preibisch et al.\ 2001; Preibisch et al.\ 2002; Ardila et al. 2000; Mart\'{\i}n et al.\ 2004; Slesnick et al.\ 2006; Slesnick et al. 2008) and the near-infrared (e.g., Lodieu et al.\ 2006; Lodieu et al. 2007). Several tens of brown dwarfs with spectral types later than M6 have been confirmed as genuine spectroscopic substellar members in UppSco (Ardila et al. 2000; Mart\'{\i}n et al.\ 2004; Slesnick et al.\ 2006; Slesnick et al. 2008; Lodieu et al.\ 2006; Lodieu et al. 2008).

Scholz et al. (2007; hereafter S07) conducted a {\it Spitzer} mid-infrared survey of 35 confirmed sub-stellar members of UppSco, selected from the works of Ardila et al. (2000) and Mart\'{\i}n et al. (2004). Among these 35 sources, 13 were found to show mid-infrared excesses explained by the presence of a circumstellar disk, resulting in a disk fraction of 37\%$\pm$9\%. Based on their $\sim$8-24$\micron$ observations, S07 noted that objects that do not show any excess emission in the mid-infrared are either diskless, or have a $\ga$5 AU inner hole in the disk. Among the 13 disk-bearing objects, only 3 sources were found to show any detectable emission in the 10 $\micron$ silicate feature. The low extent of flaring observed for these 13 sources, along with the flat silicate spectra observed for nearly 75\% of the objects thus indicated that the UppSco brown dwarfs are in a more advanced stage of disk evolution, compared to the younger Taurus objects (S07).

In this work, we consider the optical ($RI$) survey complemented by 2MASS infrared photometry, conducted by Slesnick et al.\ (2006; hereafter S06), over the full association down to $R \sim$ 20 mag (corresponding to masses of $\sim$0.02 M$_{\odot}$). This is the most complete survey available to date of the entire UppSco region down to this depth. A large number of photometric candidates were selected, and a total of 43 were confirmed as spectroscopic members, based on strong H$\alpha$ emission and weak gravity-sensitive features. Here, we present {\it Spitzer} observations for 28 UppSco brown dwarfs from the S06 survey, along with MIPS/70$\micron$ observations for the TW Hydrae Association (TWA) brown dwarf 2MASS J1139511-315921 (2M1139). We compare the structure and chemistry of the disks among brown dwarfs at ages of 5-10 Myr, considered to be the typical timescale for disk dissipation and planet formation.

\section{Observations}
\label{obs}

We have analyzed the archival {\it Spitzer}/IRAC and MIPS 24$\micron$ data (PID 20103; PI: L. Hillenbrand) for 28 UppSco brown dwarfs (spectral type $\ga$M6) from the S06 survey. S06 have identified a total of 30 brown dwarfs in their survey, though archival observations are available for 28 of these sources. Aperture photometry was performed on the post-BCD pipeline data (pipeline versions S18.7.0 for IRAC and S16.1.0 for MIPS), using the task {\it PHOT} under the IRAF package {\it APPHOT}. In the IRAC bands, the aperture radii ranged between 6 and 12 pixels, depending on the brightness of the target. The inner radius of the background annulus ranged between 3-8 pixels plus the aperture radius, while the width of the annulus ranged between 3 and 7 pixels. The inner radius and width for the background annulus was changed in order to avoid contamination from a nearby source. The zero point fluxes of 280.9, 179.7, 115.0 and 64.1 Jy were used for the IRAC channels 1 through 4, respectively. The aperture corrections as noted in the IRAC Data Handbook\footnote{Chapter 5, Table 5.7} were applied. The errors in magnitudes are between $\sim$0.02 and 0.1mag in the IRAC bands.

In the MIPS/24$\micron$ band, only four out of the 28 UppSco targets were clearly detected (detection level of $>$2$\sigma$). Aperture photometry was conducted using aperture radii of 10 pixels for the brighter objects (SCH162630 and SCH163247) and 5 pixels for the fainter ones (SCH160930 and SCH161038), with background annulus of 10-15 pixels and 5-10 pixels for the brighter and fainter sources, respectively. A zero point flux of 7.14 Jy was used, and an aperture correction of 2.05 and 2.8 was applied to the bright and faint sources, respectively. There are three sources, SCH162135, SCH162351 and SCH162354, for which only MIPS/24$\micron$ observations are available. All three of these objects were undetected at 24 $\micron$. For four sources, low-resolution {\it Spitzer}/IRS spectra are also available, and these were extracted and calibrated using the Spitzer IRS Custom Extraction (SPICE) software provided by the SSC. Further details on the processing of the IRS spectra are provided in Riaz \& Gizis (2008). For 2M1139, MIPS/70$\micron$ observations were obtained as part of the General Observer program (PID 40922; PI: Riaz). We had requested an exposure time of 10s and 20 template cycles. This brown dwarf is undetected at 70$\micron$, and we have estimated a 2$\sigma$ upper limit (Section \S$\ref{sed}$). The IRAC and MIPS photometry for all sources is listed in Table 1. The calibration uncertainty in the IRAC and MIPS bands is 2\% and 10\%, respectively.

\section{Results}
\subsection{Spectral Energy Distributions (SEDs)}
\label{sed}

Out of the 28 sources in UppSco, the SEDs for 21 objects are found to be purely photospheric, i.e. the observed emission in the IRAC and MIPS/24 $\micron$ bands coincides with the photospheric model fits within the observational errors. To fit the stellar photosphere, we have considered the NextGen models (Hauschildt et al. 1999) for solar metallicity and log {\it g} = 3.5. The effective temperature was obtained from the spectral type (SpT) using the relation for young sources in Luhman et al. (2003). The SpT for these brown dwarfs are between M6 and M8 (S06). We therefore considered $T_{eff}$ between 3000K and 2700K. The uncertainty on the temperature is of the order of $\sim$100 K. We have normalized the models to the {\it J}-band, as it best represents the photospheric emission from the central source (e.g., Cieza et al. 2005). 

\begin{figure}
 \begin{center}
    \begin{tabular}{ccc}      
    \resizebox{70mm}{!}{\includegraphics[angle=0]{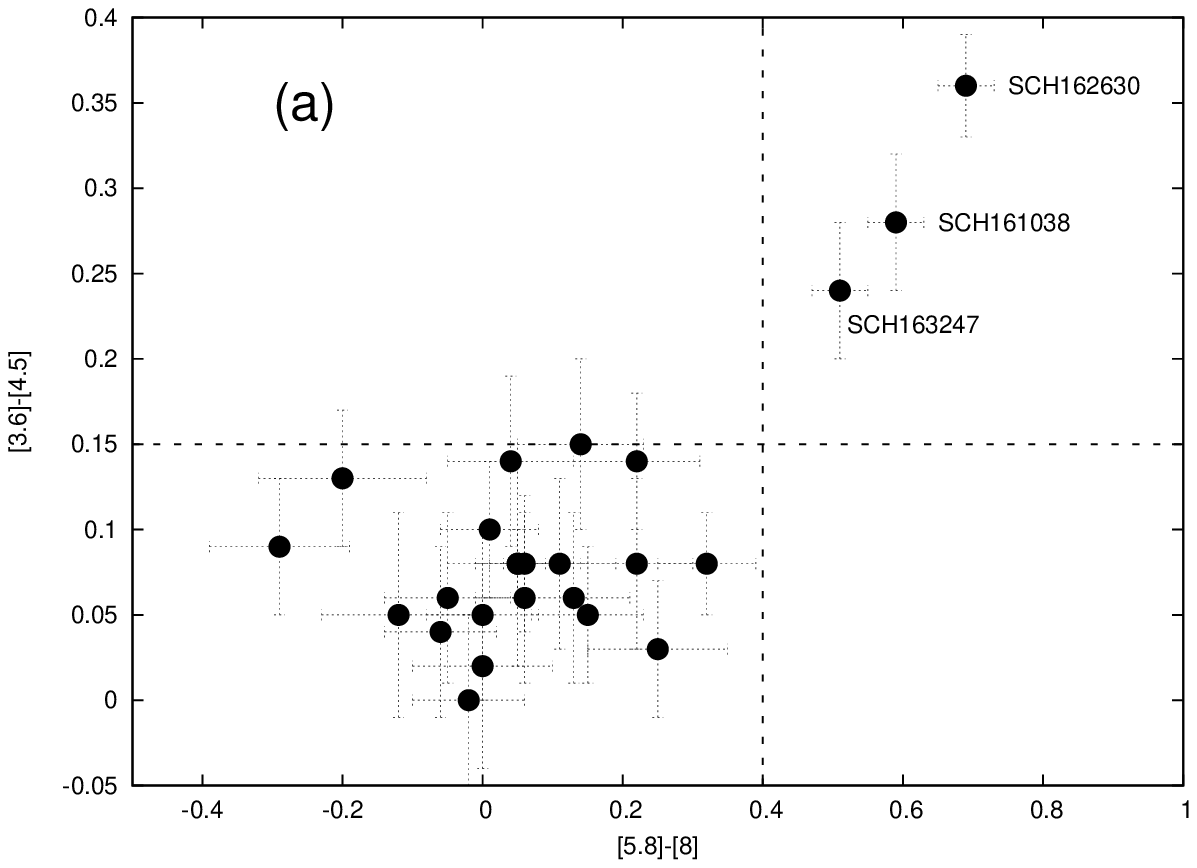}} &
     \resizebox{70mm}{!}{\includegraphics[angle=0]{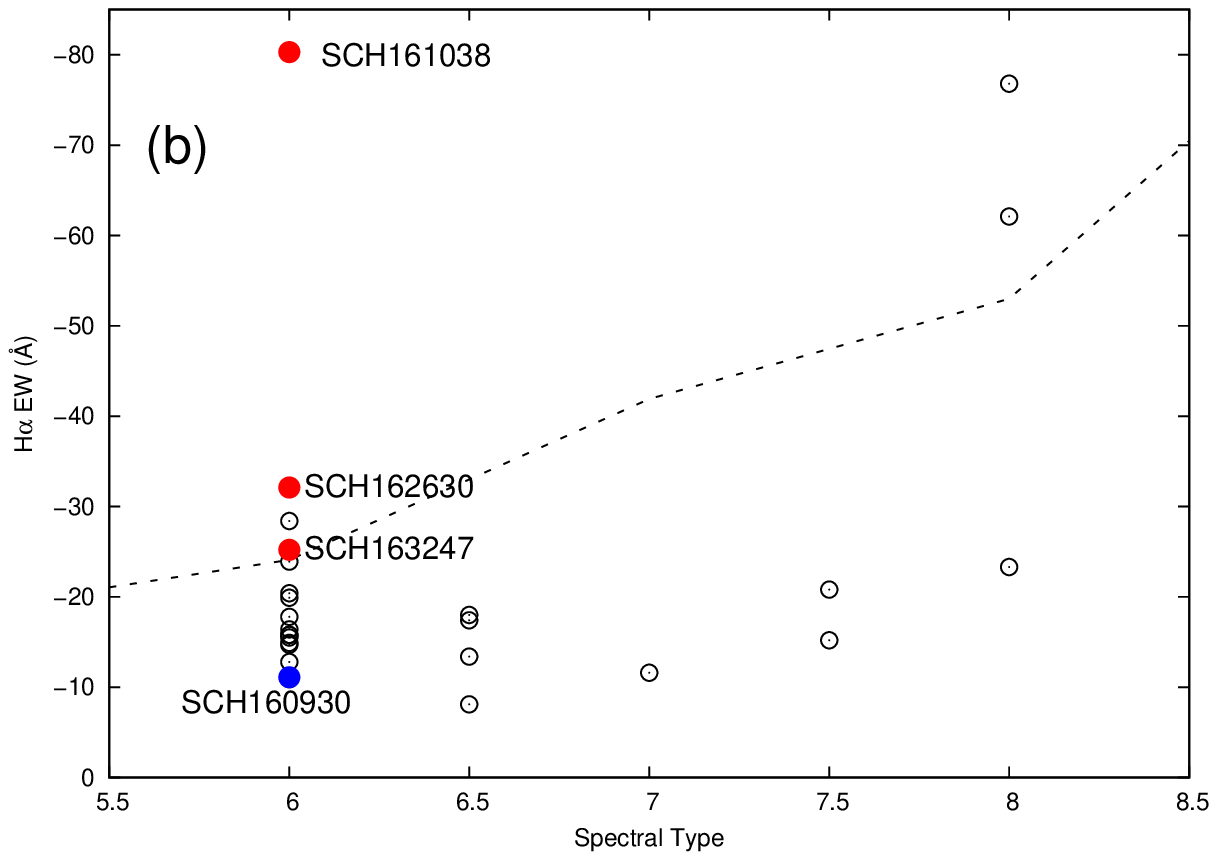}} \\   
      \resizebox{70mm}{!}{\includegraphics[angle=0]{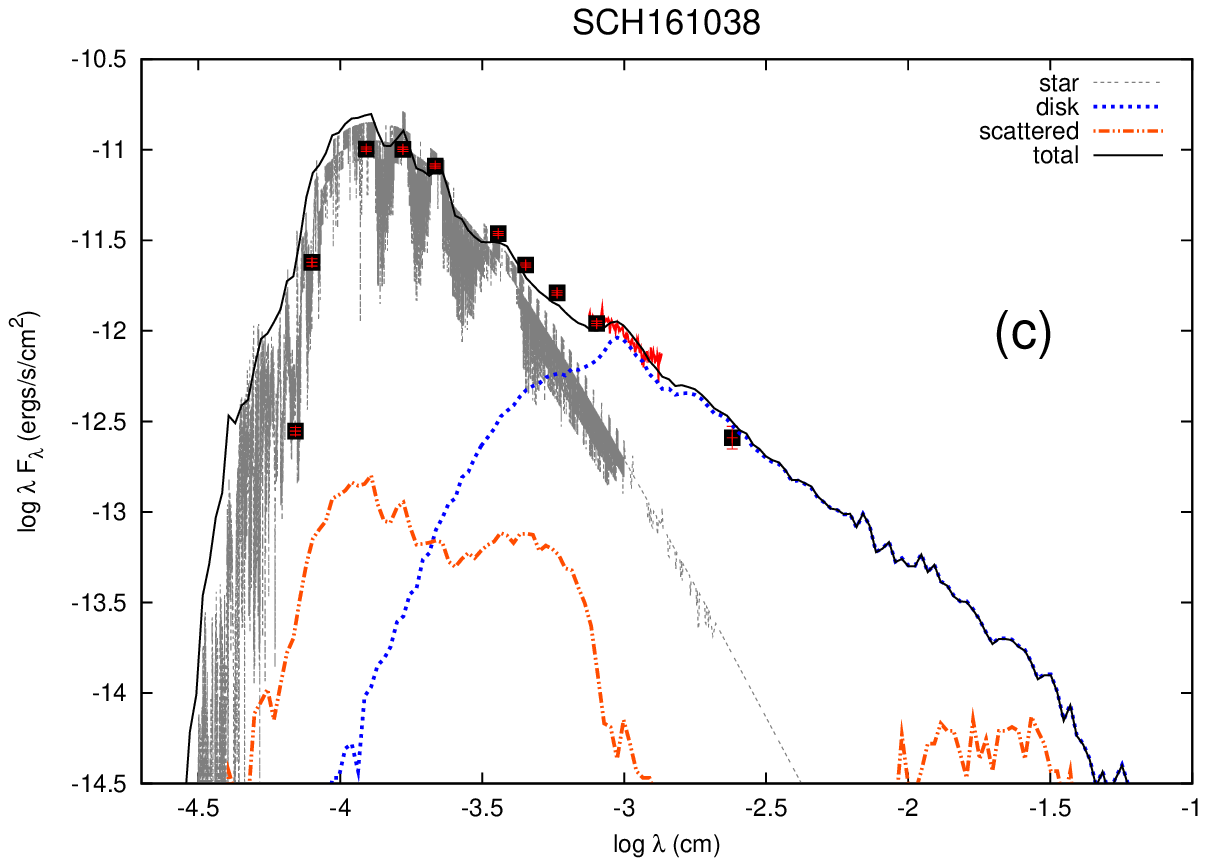}}  &   
      \resizebox{70mm}{!}{\includegraphics[angle=0]{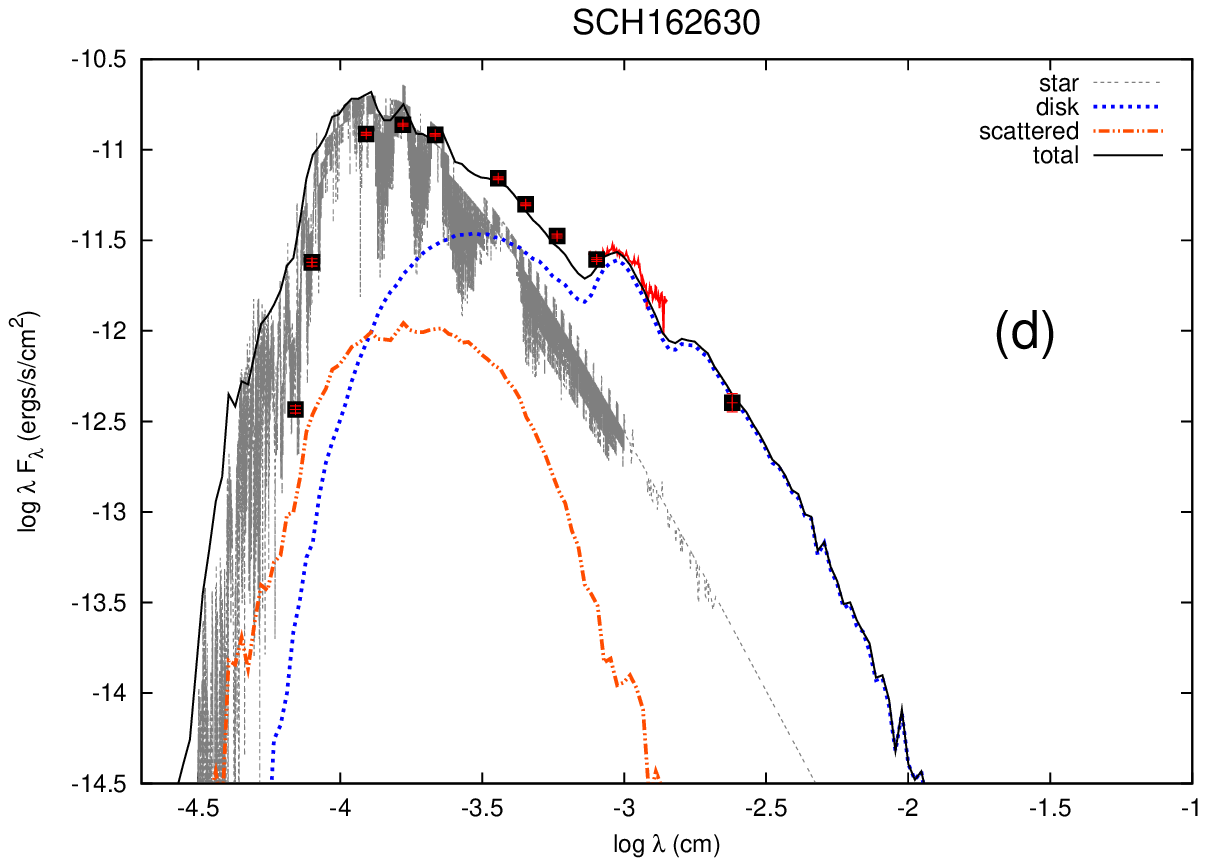}} \\
      \resizebox{70mm}{!}{\includegraphics[angle=0]{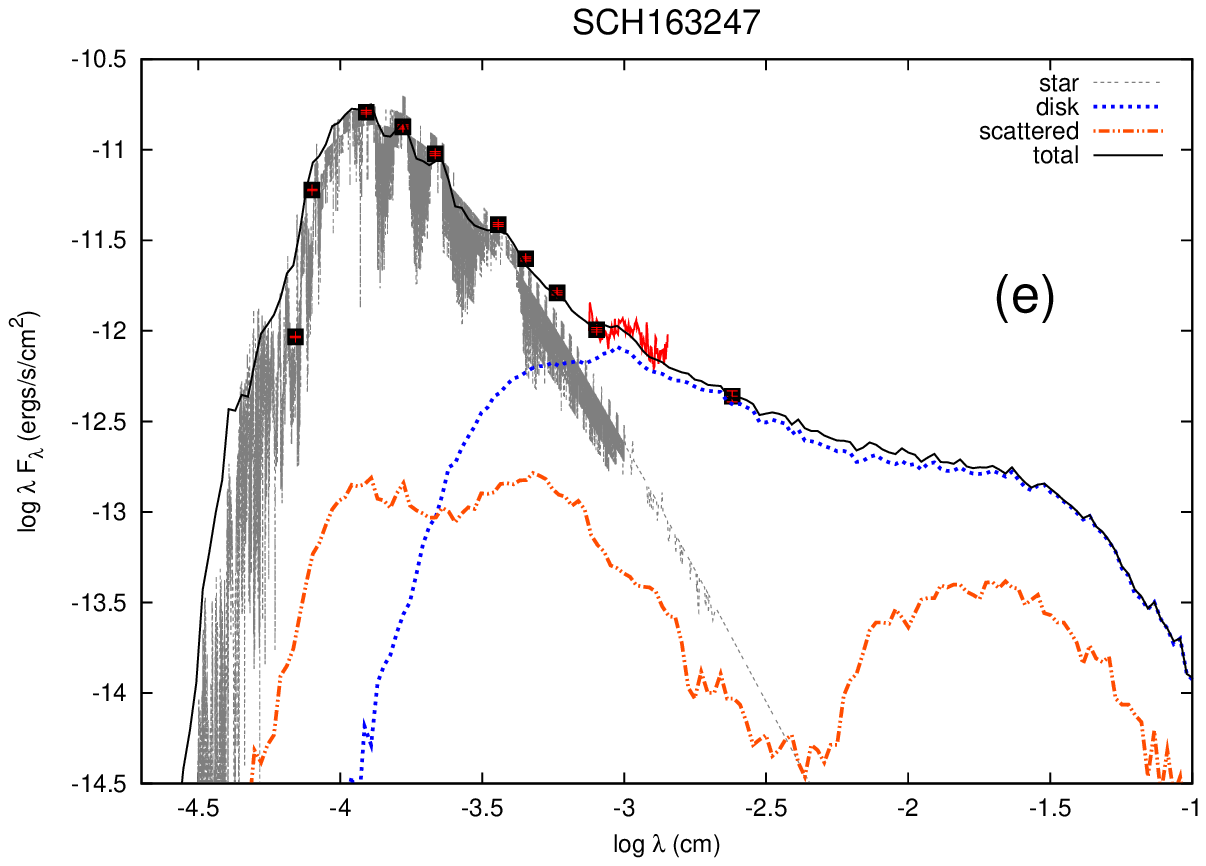}} &
      \resizebox{70mm}{!}{\includegraphics[angle=0]{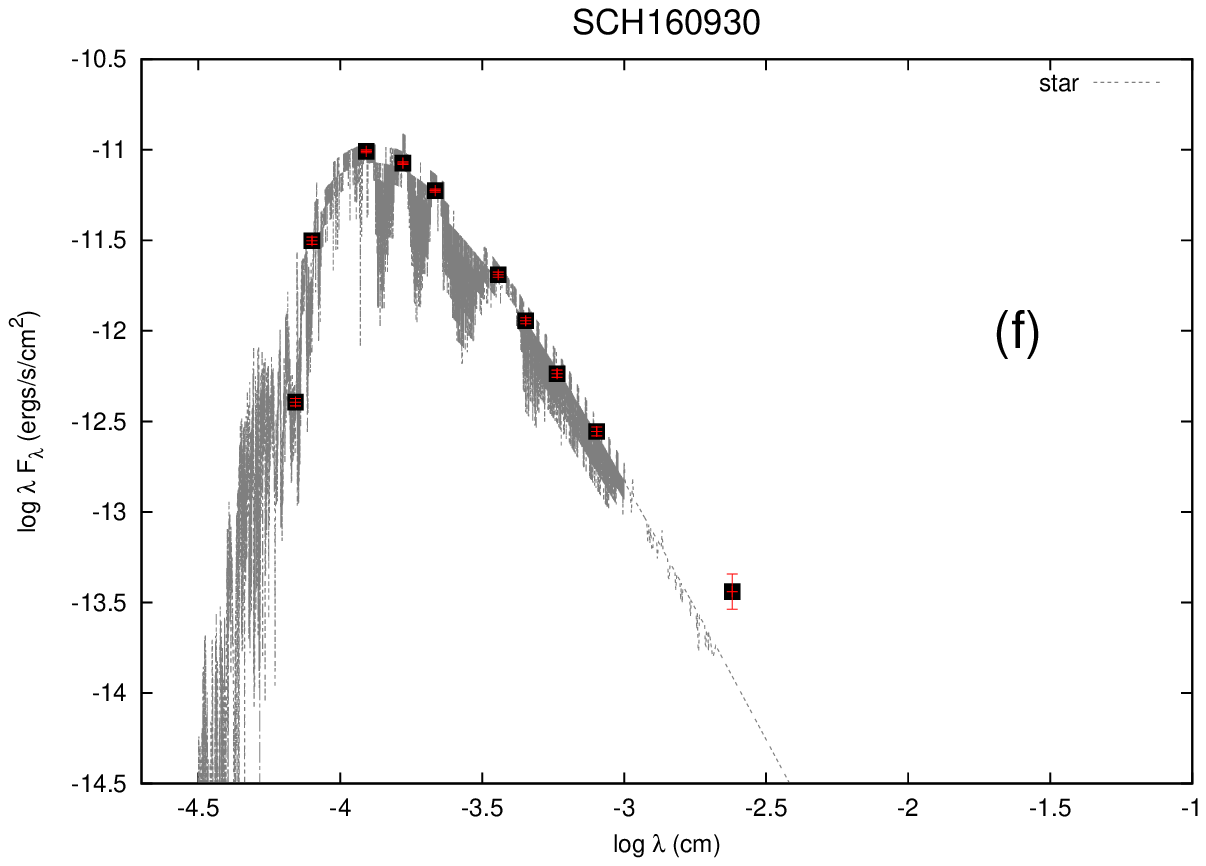}} \\
      \resizebox{70mm}{!}{\includegraphics[angle=0]{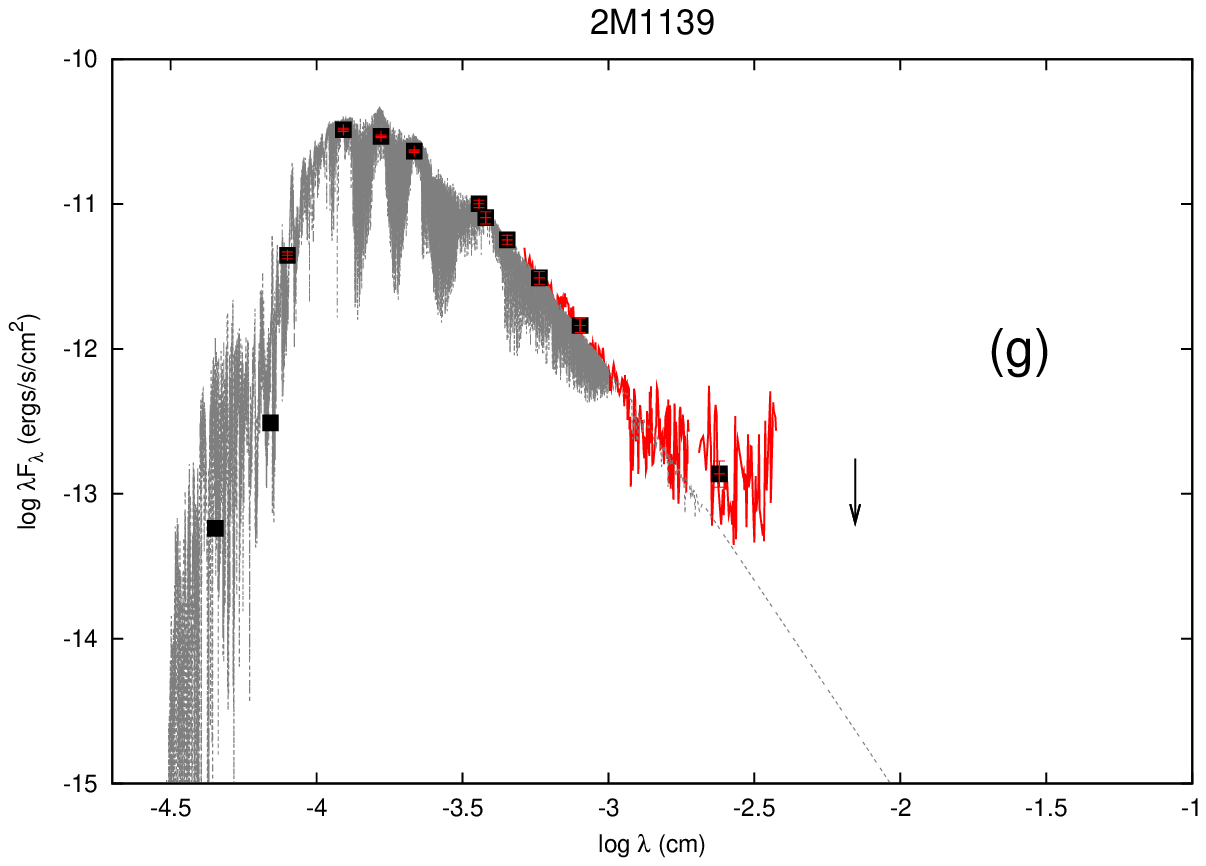}}  \\     
    \end{tabular}
    \caption{(a): IRAC ccd for 25 UppSco sources. (b): H$\alpha$ emission vs. SpT for the 28 brown dwarfs studied here. Dashed line marks the empirical accretor/non-accretor boundary. Objects with disks are denoted by red circles, while the one candidate transition disk SCH160930 is denoted by a blue circle. (c)-(e): Best model fits for the three disk-bearing objects. (f)-(g): SEDs for SCH160930 and 2M1139. }
    \label{seds}
  \end{center}
 \end{figure}

There are three other sources as noted in Section \S$\ref{obs}$ for which IRAC observations are not available, but these were undetected at 24 $\micron$. Fig.~\ref{seds}a shows the IRAC color-color diagram (ccd) for the 25 UppSco sources with IRAC observations available. We have overplotted the boundaries considered by Hartmann et al. (2005) to separate the Class II and Class III systems. There are 22 objects that lie in the lower left region characterized by nearly zero excess in both colors, and thus can be classified as Class III systems. Hartmann et al. have also found the boundary at [5.8]-[8]$\sim$0.4 to accurately separate the classical T Tauri stars (CTTS) from the weak-line T Tauri stars (WTTS). The 22 photospheric objects thus classify as WTTS, as is also evident from Fig.~\ref{seds}b where all of these objects show weak emission in the H$\alpha$ line, and lie below the empirical accretor/non-accretor boundary (S06). The empirical accretor/non-accretor boundary has been obtained from Barrado y Navascu\'{e}s \& Mart\'{i}n (2003). The SpTs and H$\alpha$ equivalent width measurements are from S06. 

In the IRAC ccd, three sources, SCH161038, SCH162630 and SCH163247, can be classified as Class II/CTTS systems. This is consistent with their H$\alpha$ emission being higher than the empirical boundary in Fig.~\ref{seds}b, at a SpT of M6. The SEDs for these brown dwarfs (Fig.~\ref{seds} c-e) show excess emission in at least two of the IRAC bands, and at 24$\micron$. One object, SCH160930, shows a small excess at 24$\micron$, with a $F_{obs}/F_{phot}$ of 2.88 (Fig.~\ref{seds}f). The emission at shorter wavelengths, however, is photospheric. This therefore may be a candidate transition disk, i.e. a system with an inner hole in the disk. The inner disk region may be either devoid of any disk material or may contain optically thin dust. SCH160930 is classified as a Class III/WTTS source, based on its IRAC colors and weak H$\alpha$ emission. Longer wavelength observations are required to confirm the presence of a transition disk around this brown dwarf.

We have also considered the classification scheme by Lada et al. (2006), which is based on the 3.6-8 $\micron$ slope, $\alpha$, of the IRAC SED and classifies objects into optically thick disks, anemic (transition) disks or photospheric sources. Sources with optically thick disks are characterized by $\alpha >$ -1.80, diskless sources by $\alpha <$ -2.56, while sources with -2.56 $< \alpha <$ -1.80 are considered anemic disks, i.e. objects with optically thin inner disks or disks with inner holes. For the three confirmed disk-bearing objects SCH161038, SCH162630 and SCH163247, we find $\alpha$ of -1.52, -1.31 and -1.67, respectively, thus classifying all three to possess optically thick disks. The values of $\alpha$ for SCH161038 and SCH163247 are though close to the optically thick/anemic disk boundary, which can be explained by the weak or no excess emission observed for these sources shortward of $\sim$5 $\micron$. Based on their $\sim$8-24 $\micron$ observations, S07 had noted that objects that do not show any excess emission in the mid-infrared are either diskless, or have a $\ga$5 AU inner hole in the disk. They however could not constrain inner holes of sizes $\la$1 AU due to the absence of observations in the $\sim$3-8$\micron$ wavelength range. For the three sources studied here that show excess emission in the IRAC bands, we estimate inner disk holes of $\ga$0.1 AU (Section \S$\ref{diskmodel}$). For the candidate transition disk SCH160930, the inner disk radius could be as large as $\sim$1 AU. We find an $\alpha$ of -2.6 for this source, consistent with it being photospheric in the IRAC bands. 

For the case of the TWA brown dwarf 2M1139, Riaz et al. (2006) had reported an excess emission at 24$\micron$ with a $F_{obs}/F_{phot}$ of $\sim$3, while Riaz \& Gizis (2008) had noted an increase in the flux densities in the IRS spectrum for this object, longward of 24$\micron$. The IRS spectrum though was at a very low S/N ($<$5). Based on the same observations, Morrow et al. (2008) had reported this object to be photospheric upto wavelengths of $\sim$20 $\micron$. We have estimated a 2$\sigma$ upper limit at 70 $\micron$ for 2M1139 by measuring the flux in the regions that lie close to the position of the source, but which do not contain any other sources. The measured flux thus represents the noise level in that region (e.g., S07). Though the 2$\sigma$ upper limit for 2M1139 is well in excess of the predicted photospheric emission ($F_{obs}/F_{phot} \sim$ 70) (Fig.~\ref{seds}g), this brown dwarf is undetected at 70 $\micron$, and therefore our observations are inconclusive regarding the presence or absence of a disk around this source.

\subsection{Disk Modeling}
\label{diskmodel}

We have used the 2-D radiative transfer code by Whitney et al. (2003) to model the disks around SCH161038, SCH162630 and SCH163247. The circumstellar geometry consists of a rotationally flattened infalling envelope, bipolar cavities, and a flared accretion disk in hydrostatic equilibrium. The disk density is proportional to $\varpi^{-\alpha}$, where $\varpi$ is the radial coordinate in the disk midplane, and $\alpha$ is the radial density exponent. The disk scale height increases with radius, $h=h_{0}(\varpi / R_{*})^{\beta}$, where $h_{0}$ is the scale height at $R_{*}$ and $\beta$ is the flaring power. For the stellar parameters, we have used a $T_{eff}$=3000K, $M_{*}$=0.06 $M_{\sun}$ and $R_{*}$=0.6 $R_{\sun}$. The effective temperature was obtained from the $T_{eff}$-SpT relation in Luhman et al. (2003), while the mass and radius were obtained by considering the 5 Myr isochrone from Baraffe et al. (2003) for a $T_{eff}$ of 3000K. The NextGen (Hauschildt et al. 1999) atmosphere file for a $T_{eff}$ of 3000K, and log {\it g} = 3.5 was used to fit the atmosphere spectrum of the central sub-stellar source. A distance of 145$\pm$2 pc (de Bruijne et al.\ 1997) was used to scale the output fluxes from the models to the luminosity and distance of the three brown dwarfs. 

Figs. 1c-e show the model fits obtained for SCH161038, SCH162630 and SCH163247. Also indicated are the separate contributions from the disk, the stellar photosphere and the scattered flux. Detailed discussions on the fitting procedure and the variations in the model SEDs with the different disk parameters are provided in Whitney et al. (2003) and Riaz \& Gizis (2007). Here we provide a brief description of the best model-fits (based on the lowest reduced-$\chi^{2}$ value) obtained for these brown dwarfs. The models used here consider three grain models: large grains with a size distribution that decays exponentially for sizes larger than 50 $\micron$ up to 1 mm, grains of sizes with $a_{max}\sim$ 1 $\micron$, and ISM-like grains with $a_{max} \sim$ 0.25 $\micron$. We have varied these grain models in the disk midplane and the upper atmosphere. The object SCH162630 shows more emission in the 10 $\micron$ silicate feature compared to the other two sources (Fig.~\ref{silicate}f). For this source, sub-micron sized grains were placed in both the midplane and the disk surface layers. This produces a model SED with a more peaked emission at 10$\micron$ and a steeper slope at far-infrared/sub-mm wavelengths (e.g., Riaz \& Gizis 2007). SCH163247 shows weaker emission while the silicate feature is nearly flat for SCH161038 (Fig.~\ref{silicate}d and g). For both of these, the large grain model with $a_{max}$ of 1 mm was considered in the disk midplane, while $a_{max} \sim$ 1 $\micron$ grains were placed in the upper layers. This results in a weaker silicate emission feature and a flatter sub-mm slope in the model SED. For SCH161038, using the large-grain model in the disk atmosphere, which results in a flat silicate feature, also provides a good fit to the observed SED. Considering that we do not have any far-infrared/sub-mm observations for these objects, the 10 $\micron$ feature provides the only way to constrain the different grain sizes considered by these disk models. A more detailed compositional analysis of the silicate features is provided in Section \S$\ref{composition}$. 

As mentioned, the disk scale height varies as $h=h_{0}(\varpi / R_{*})^{\beta}$, where $h_{0}$ is the scale height at $R_{*}$ and $\beta$ is the flaring power. We varied both $h_{0}$ and $\beta$ to determine the amount of flaring in these disks. The variations in the model SEDs as $\beta$ is lowered are more evident for $\lambda$ $>$ 10 $\micron$, and the 24 $\micron$ observation provides a good constraint to the flaring power. The model SED is highly flared for $\beta$=1.25, which is considered a typical value for models of T Tauri disks in hydrostatic equilibrium (e.g., Walker et al. 2004), while the structure is flat for $\beta$=1.0 or smaller values (e.g., Riaz \& Gizis 2007). The 24$\micron$ emission for SCH163247 is higher than the other two objects, and a slightly higher flaring power of 1.17 provides a good fit to the observed SED. In comparison, SCH161038 and SCH162630 show flatter structures, with $\beta$ of 1.1 and 1.0 providing good fits, respectively. The scale height at $R_{*}$ was set to a small value of 0.01 for SCH161038 and SCH163247, while a higher value of 2.8 was required to fit the stronger emission observed near 10 $\micron$ and at shorter wavelengths for SCH162630. None of the sources therefore show evidence for strong flaring in the disk, as indicated by the low values of $\beta$ between 1.0 and 1.17. This is consistent with the findings of S07 who reported a low degree of flaring in the 13 UppSco brown dwarf disks studied in their survey. 

We have explored a range of inclination angles to the line of sight. Due to binning of photons in the models, there are a total of 10 viewing angles, with face-on covering 0-18$\degr$ inclinations. The mid- and far-IR fluxes increase with decreasing inclinations, as the emission at these wavelengths is from an optically thick disk, while the optically thin millimeter fluxes are independent of the inclination (e.g., Walker et al. 2004). An inclination that is more face-on ($<$40$\degr$) provides a good fit to all three objects, with the bin of 18$\degr$-32$\degr$ providing the best-fit. The disk outer radius was fixed at 100 AU, since variations in this parameters are more evident at far-infrared/sub-mm wavelengths (e.g., Riaz \& Gizis 2007). As mentioned in Section \S$\ref{sed}$, SCH161038 and SCH163247 show photospheric emission upto wavelengths of $\sim$5$\micron$, indicating inner holes in the disks. For these two sources, an inner disk radius of 3$R_{sub}$ was considered, where $R_{sub}$ is the dust sublimation radius (1$R_{sub}$ = 3.5$R_{*}$). A 3$R_{sub}$ radius thus corresponds to an inner hole of $\sim$0.2 AU in the disk. For the third source SCH162630, we considered an inner disk radius of 1 $R_{sub}$, equal to about 0.06 AU. Thus for the three sources studied here that show excess emission in the IRAC bands, we estimate inner disk holes of $\ga$0.1 AU. The disk mass for modeling was set to 1E-4$M_{\sun}$; this is the typical disk mass obtained by Scholz et al. (2006) through modeling the 1.3mm observations for a sample of Taurus brown dwarfs. For the case of SCH162630, however, a steeper slope is observed between the red edge of the IRS spectrum and the 24$\micron$ point (Fig.~\ref{seds}d). Reducing the disk mass to 3E-7$M_{\sun}$ provides a better fit to the observed SED for this source. However in the absence of millimeter data, it is impossible to constrain the model SEDs and obtain a good estimate on the disk mass. 

There are degeneracies in the model fits presented here, considering that we have only the IRAC and MIPS/24 $\micron$ data points to model these disks. The uncertainties in the stellar parameters could also result in changes in the disk parameters. There are mainly six free parameters related to the disk emission in these models, that can be varied to obtain a good fit. We had fixed the disk mass and the outer disk radius given the absence of longer wavelength data. There are no measurements available for the disk mass accretion rates for these sources.  We had fixed this parameter at a value of 10$^{-10} M_{\sun} yr^{-1}$, a typical value observed among accreting T Tauri stars (e.g., Muzerolle et al. 2003), considering that these sources are more likely to be accreting (Fig.~\ref{seds}b). However, it is possible to constrain to some extent the amount of flaring in the disk, the inclination angle as well as the inner disk radius, given the data at hand. A wide spread is observed at 24 $\micron$ in the model SEDs for the different flaring powers, and variations in the model SEDs with the disk inclination angles are more evident at wavelengths between $\sim$8 and 60 $\micron$, while the optically thin millimeter fluxes are independent of the inclination (e.g., Riaz \& Gizis 2007). As mentioned above, inclinations less than $\sim$40$\degr$ provide better fits to all three sources. We can estimate a range in flaring power of 1.0-1.1 for SCH161038 and SCH162630, while a slightly higher range of 1.15-1.18 provides a good fit to SCH163247. The inner disk radii can be estimated between $\sim$0.1-0.3 AU for all three sources, although a radius of  7$R_{sub}$ ($\sim$24$R_{*}$$\approx$0.6AU) results in higher fluxes near the 10 $\micron$ silicate band and at longer wavelengths, while the emission around 3 $\micron$ decreases. This large a radius is not a good fit to the observed SEDs. It is thus possible to obtain some estimate on the extent of inner disk clearing in the disk with the observations available.

\subsection{The 10$\micron$ Silicate Emission Feature}
\label{composition}

\begin{figure}
 \begin{center}
    \begin{tabular}{ccc}      
      \resizebox{55mm}{!}{\includegraphics[angle=0]{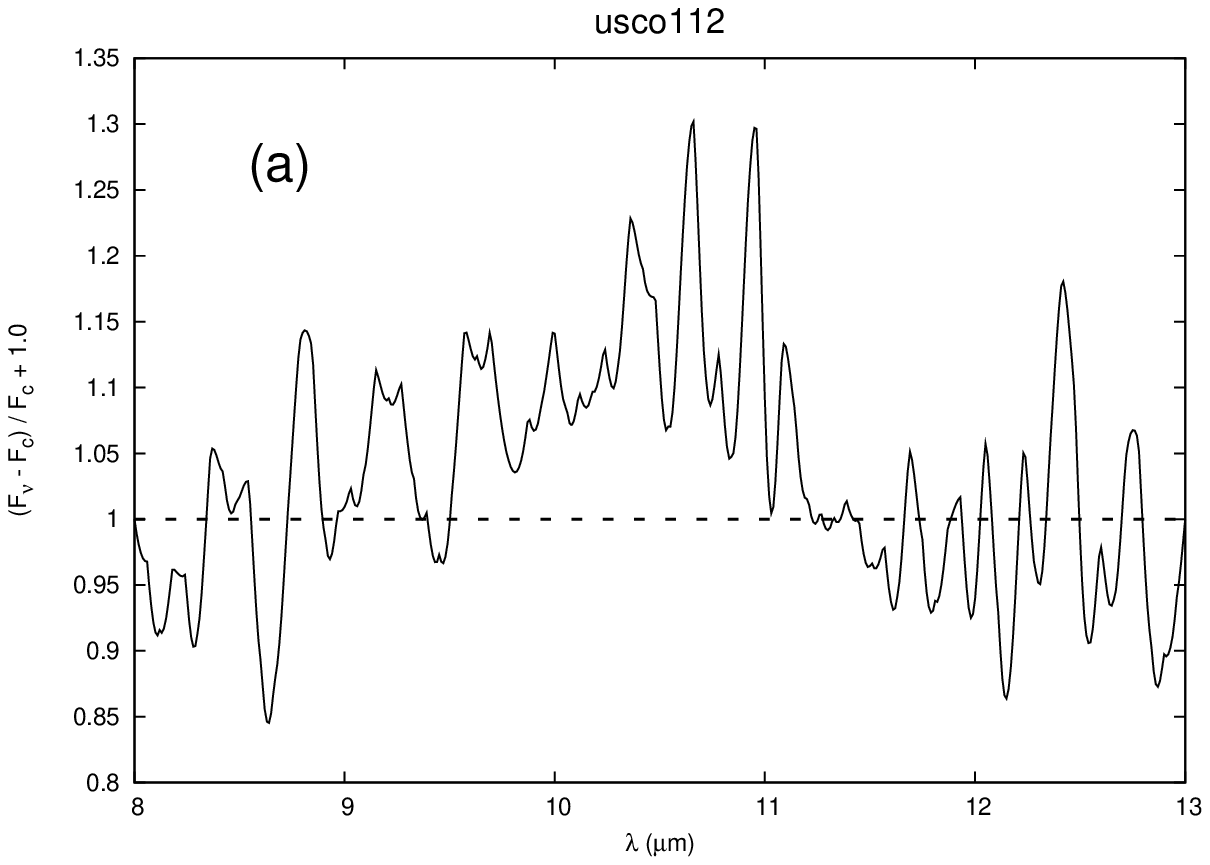}} &  
      \resizebox{55mm}{!}{\includegraphics[angle=0]{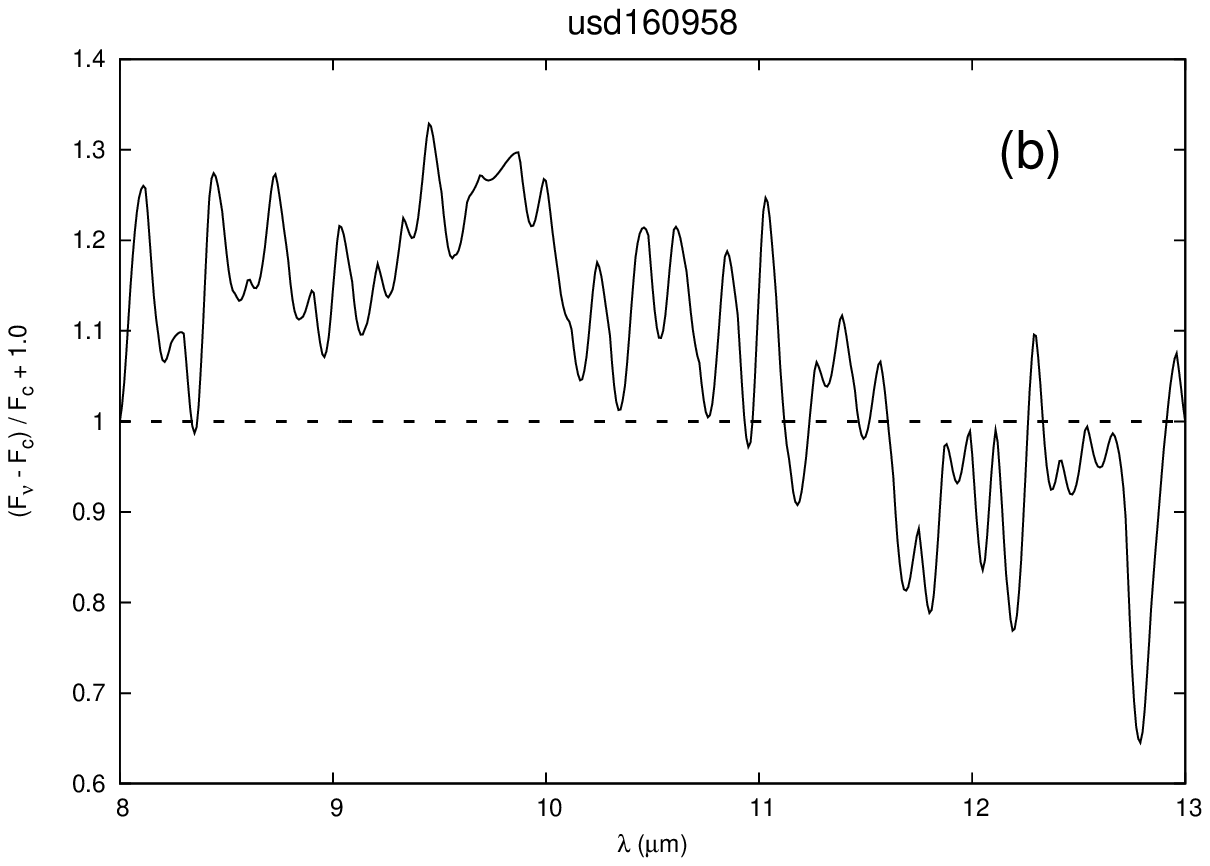}}  \\   
      \resizebox{55mm}{!}{\includegraphics[angle=0]{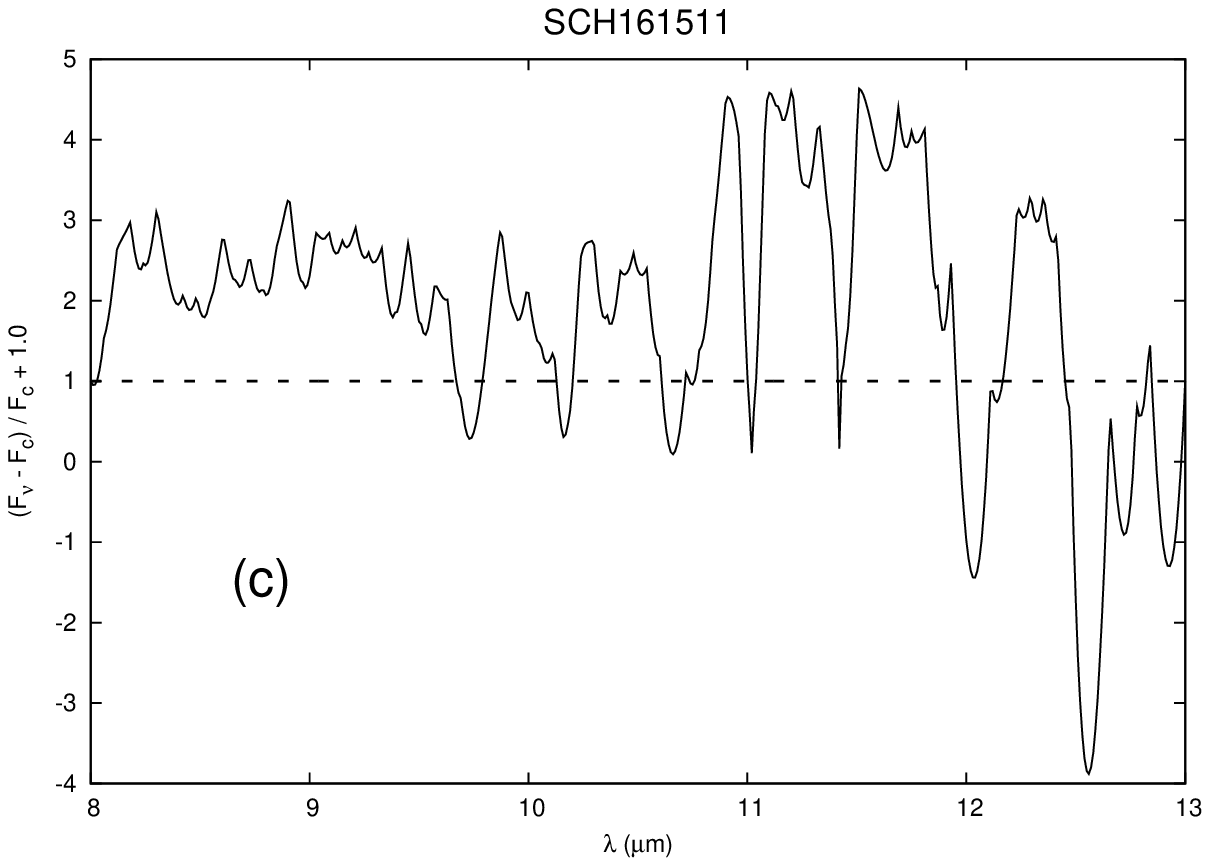}} &
      \resizebox{55mm}{!}{\includegraphics[angle=0]{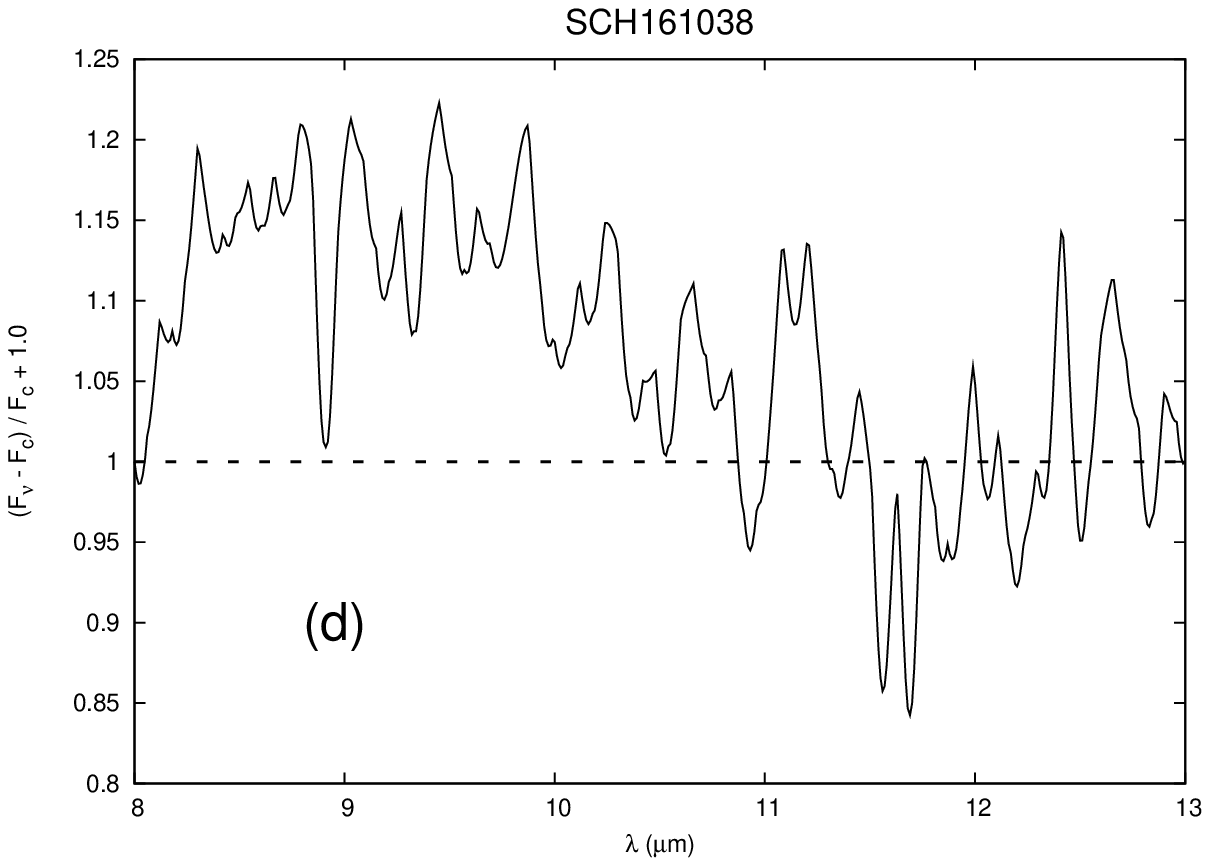}} \\   
    \end{tabular}
    \caption{The normalized continuum-subtracted spectra in units of ($F_{\nu} - F_{c})/F_{c}$ for the weak/flat features. Dashed line represents the continuum. }
    \label{silicate}
  \end{center}
 \end{figure}

In Riaz (2009; hereafter R09), we had presented compositional fits to the 10$\micron$ silicate emission features for three UppSco brown dwarfs: UScoCTIO J160026.6-205632.0 (usco112), UScoCTIO J160958.5-234518.6 (usd160958) and UScoCTIO J161939.8-214535.1 (usd161939). The coordinates for these sources are from Ardila et al. (2000). To obtain a compositional fit, five dust species were considered: amorphous olivine and pyroxene, crystalline enstatite, crystalline forsterite, and silica. For the amorphous species, two grain sizes of 0.1 and 2 $\micron$ were considered, while only the sub-micron size was considered for the crystalline silicates. A model spectrum was then constructed by calculating the sum of the emission from a featureless power-law continuum, and the emission from the different dust species (at the two different grain sizes). To fit the model spectrum, the $\chi^{2}$-minimization method as outined in van Boekel et al. (2005) was used. A detailed discussion on the method and analysis is provided in R09. The uncertainties for the mass fractions were obtained by taking into consideration both the noise in the spectrum, and the errors for the observed flux densities. In addition to a model that included all of the five dust species, R09 had constructed two more models: model `(a)' that did not consider any crystalline silicates, and model `(b)' that included large amorphous grains only. The two additional models respectively highlighted the importance of the crystallization and the grain growth processes in the disk. For the sources usco112 and usd160958, the silicate emission features are very weak and show a low contrast above the continuum (Fig.~\ref{silicate}a-b). R09 had found all three models to provide a good fit to the observed spectra for these sources, with a difference in the reduced-$\chi^{2}$ value between the best-fit and these additional fits of $<$0.1. This thus indicated the large uncertainties in their model-fits and the derived mass fractions. Among the S06 brown dwarfs, there are two similar cases of SCH161511 and SCH161038 that show weak, low-contrast features. The flat feature observed for SCH161511 is consistent with it being a photospheric source. Fig.~\ref{silicate} shows the normalized continuum-subtracted spectra for these four objects with weak features. In order to normalize the silicate feature to the continuum, we first approximated the underlying continuum by connecting the two end points of the spectrum, i.e. the observed fluxes at 8 and 13 $\micron$ (averaging over 0.2 $\micron$ at the end points; Kessler-Silacci et al. 2005). The normalized silicate spectrum is then defined in units of [($F_{\nu}-F_{c}$)/$F_{c}$], where $F_{\nu}$ and $F_{c}$ are the observed and the continuum normalized fluxes, respectively. 

\begin{figure}
 \begin{center}
    \begin{tabular}{ccc}      
    \resizebox{50mm}{!}{\includegraphics[angle=0]{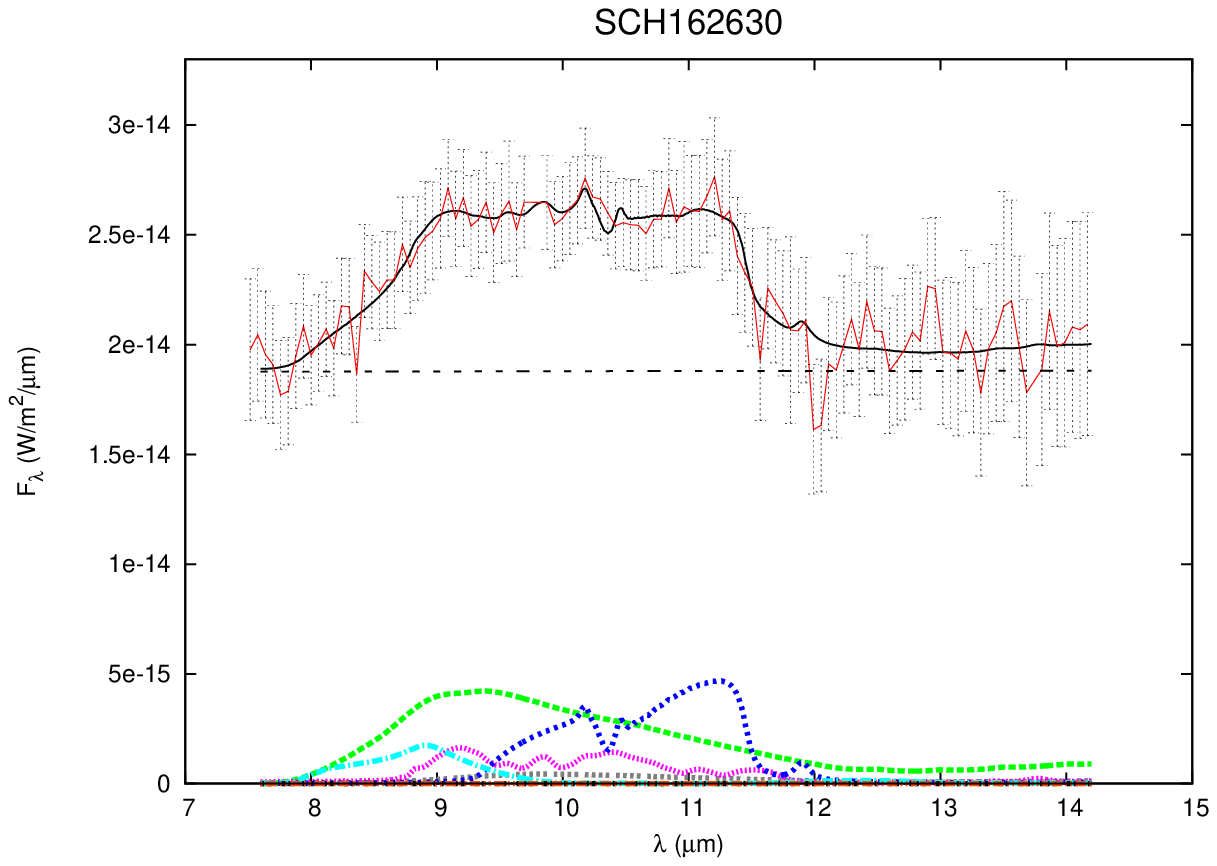}} &  
      \resizebox{50mm}{!}{\includegraphics[angle=0]{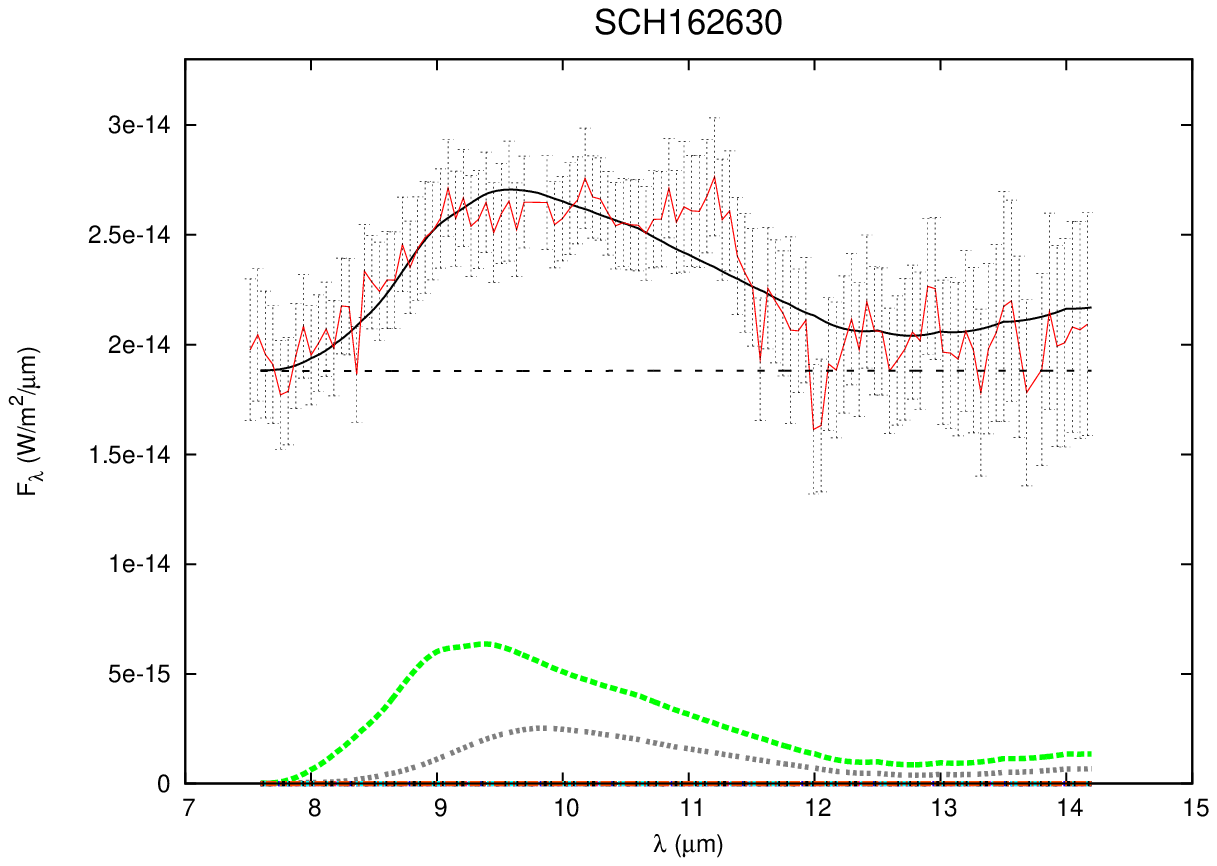}}  &  
      \resizebox{50mm}{!}{\includegraphics[angle=0]{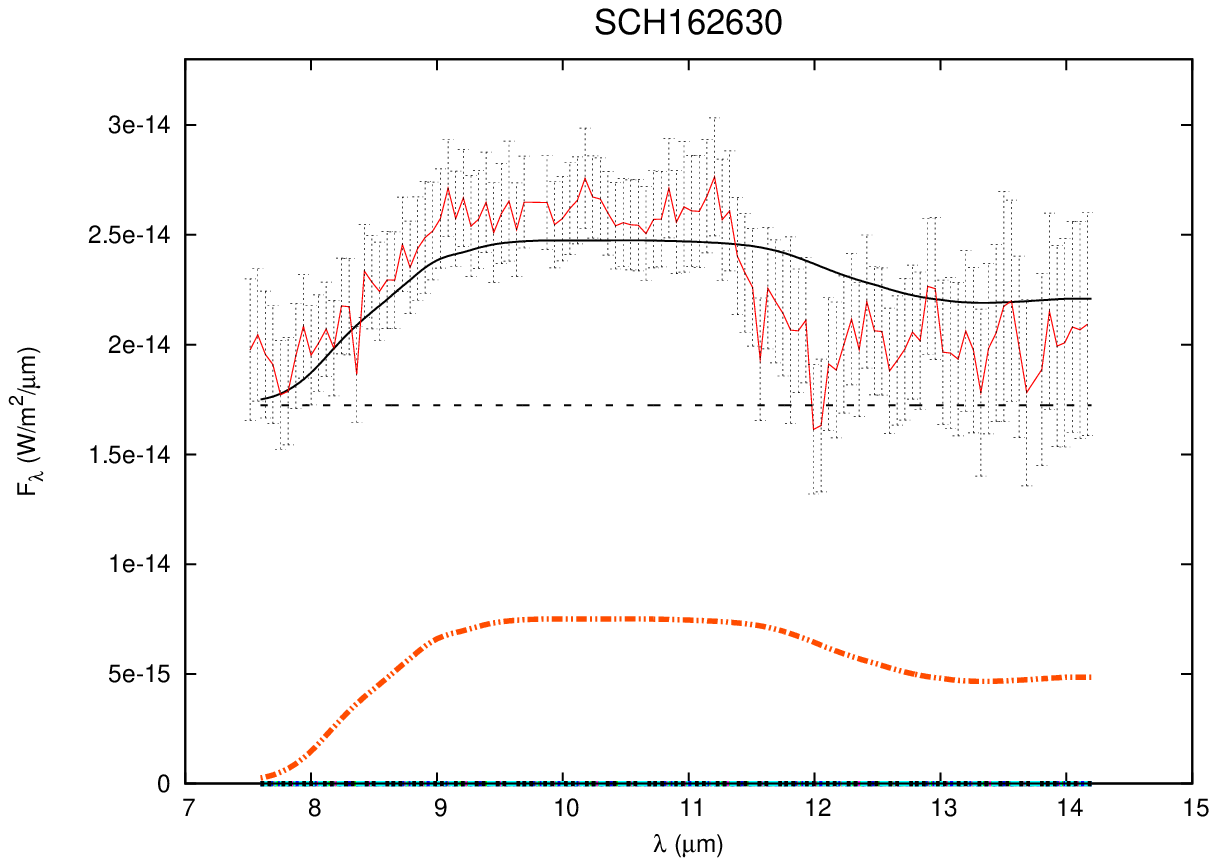}} \\
      \resizebox{50mm}{!}{\includegraphics[angle=0]{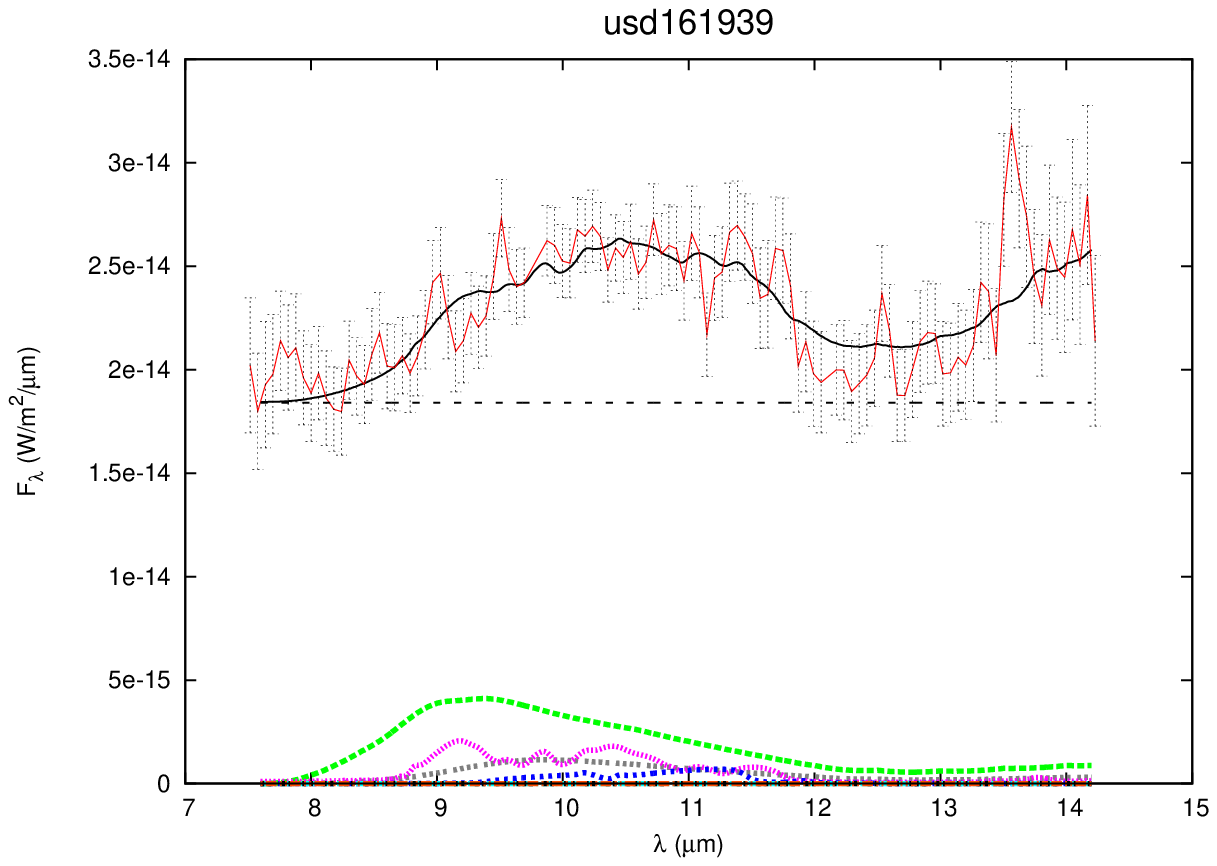}} &
      \resizebox{50mm}{!}{\includegraphics[angle=0]{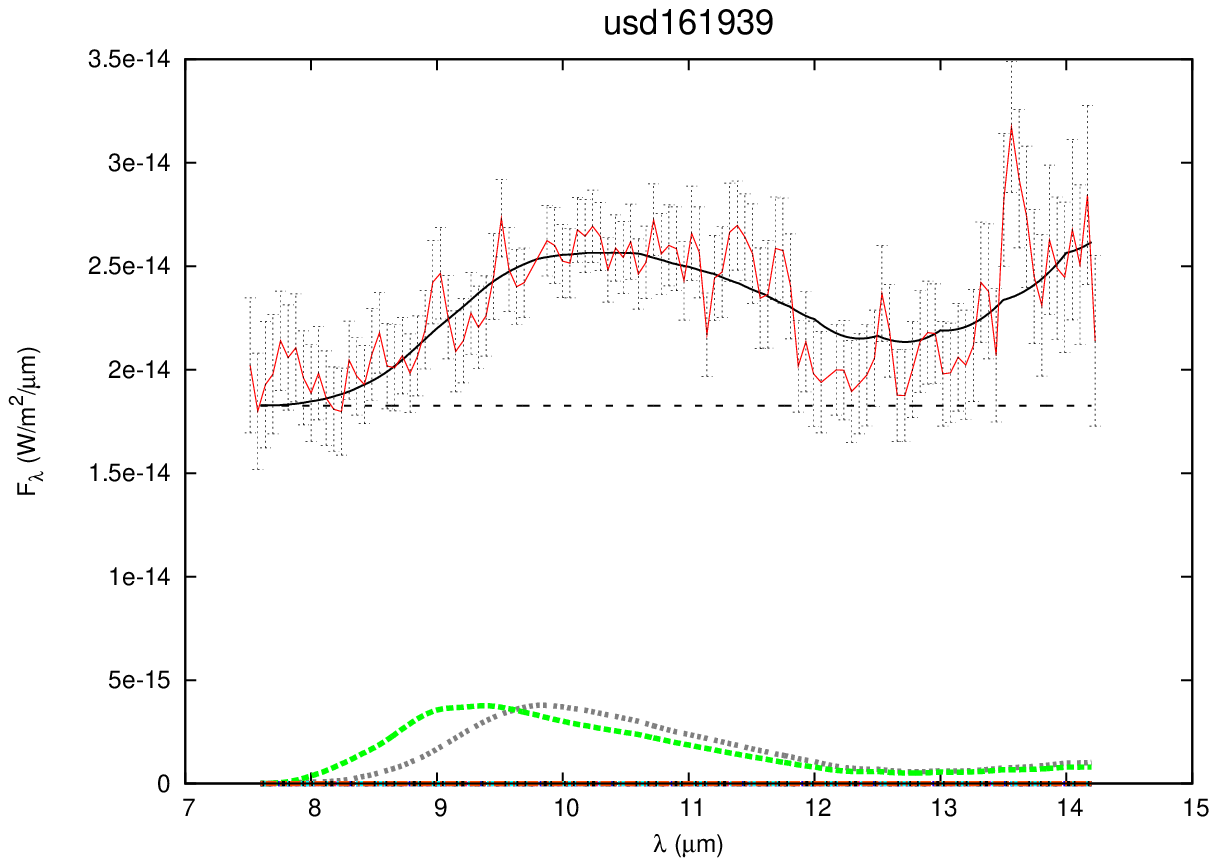}}  &  
      \resizebox{50mm}{!}{\includegraphics[angle=0]{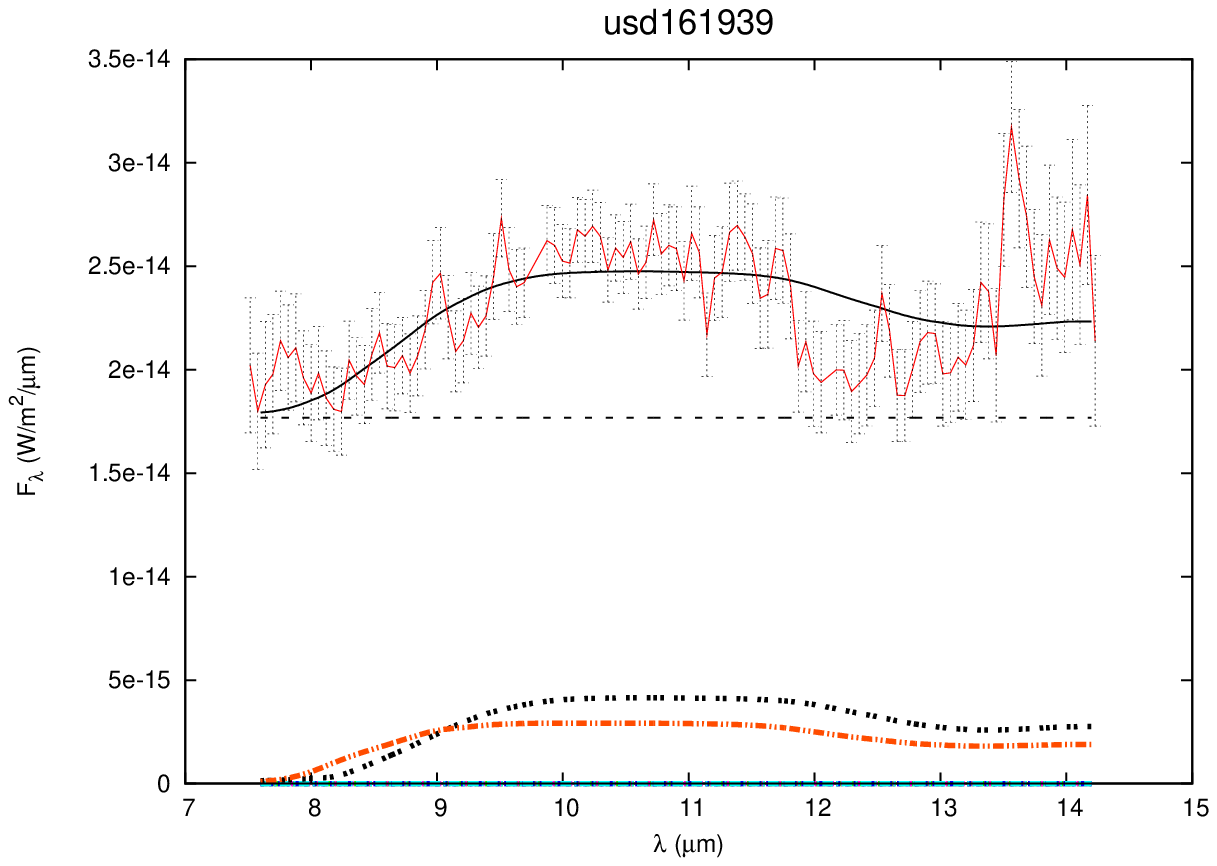}} \\
      \resizebox{50mm}{!}{\includegraphics[angle=0]{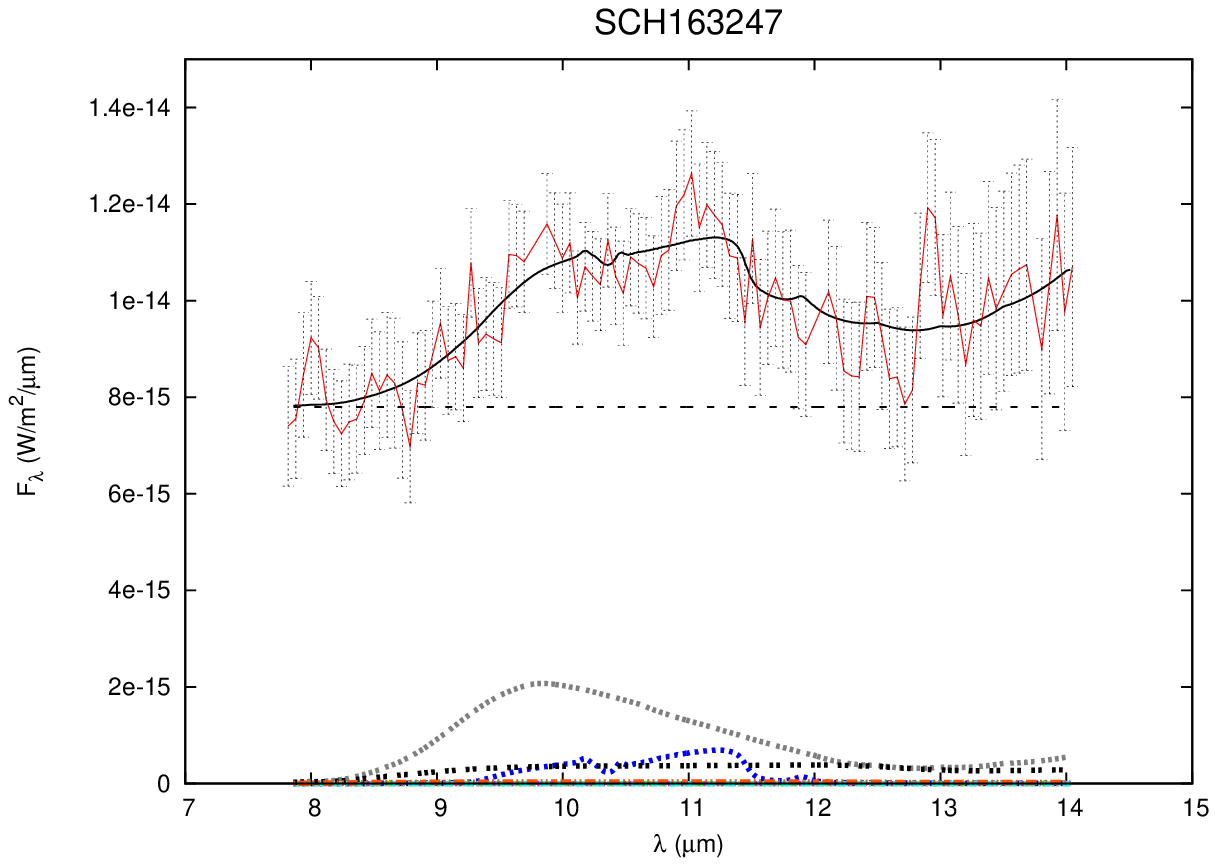}} &     
       \resizebox{50mm}{!}{\includegraphics[angle=0]{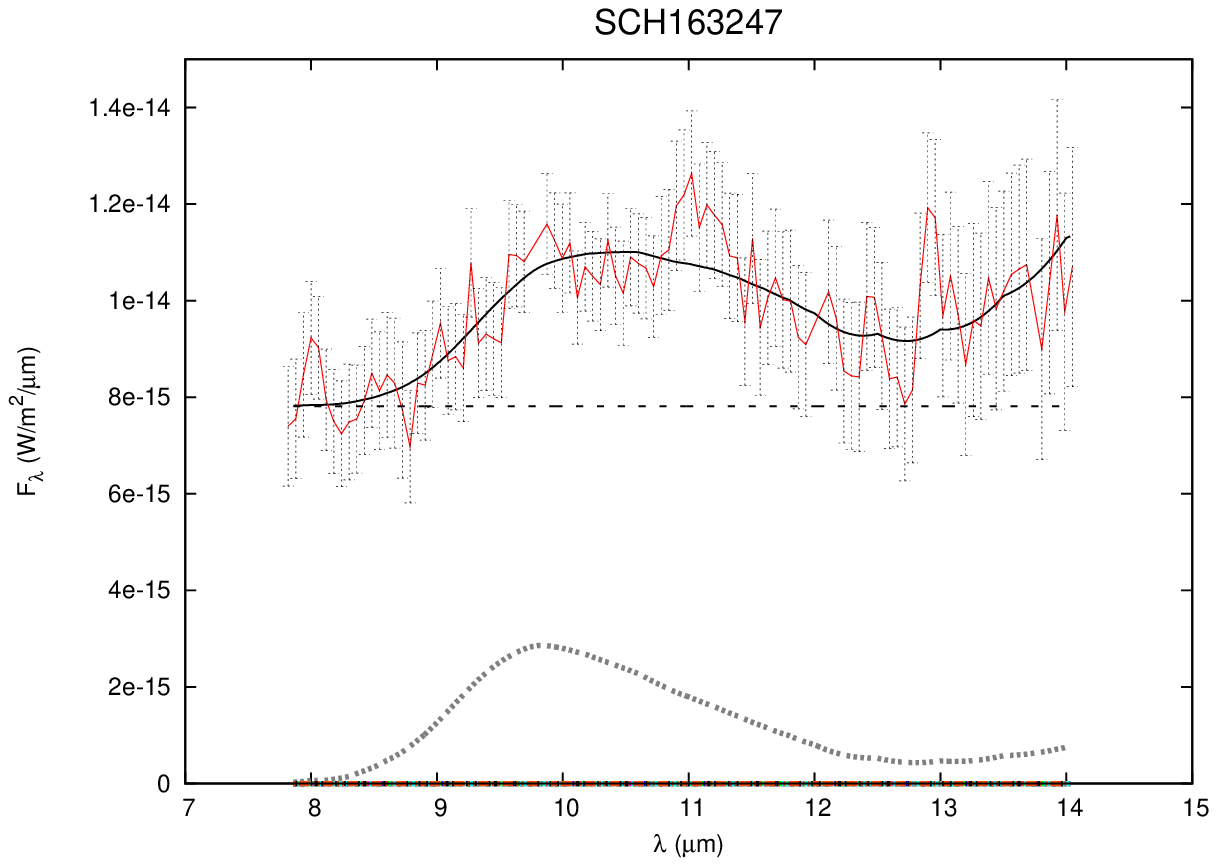}}  &  
      \resizebox{50mm}{!}{\includegraphics[angle=0]{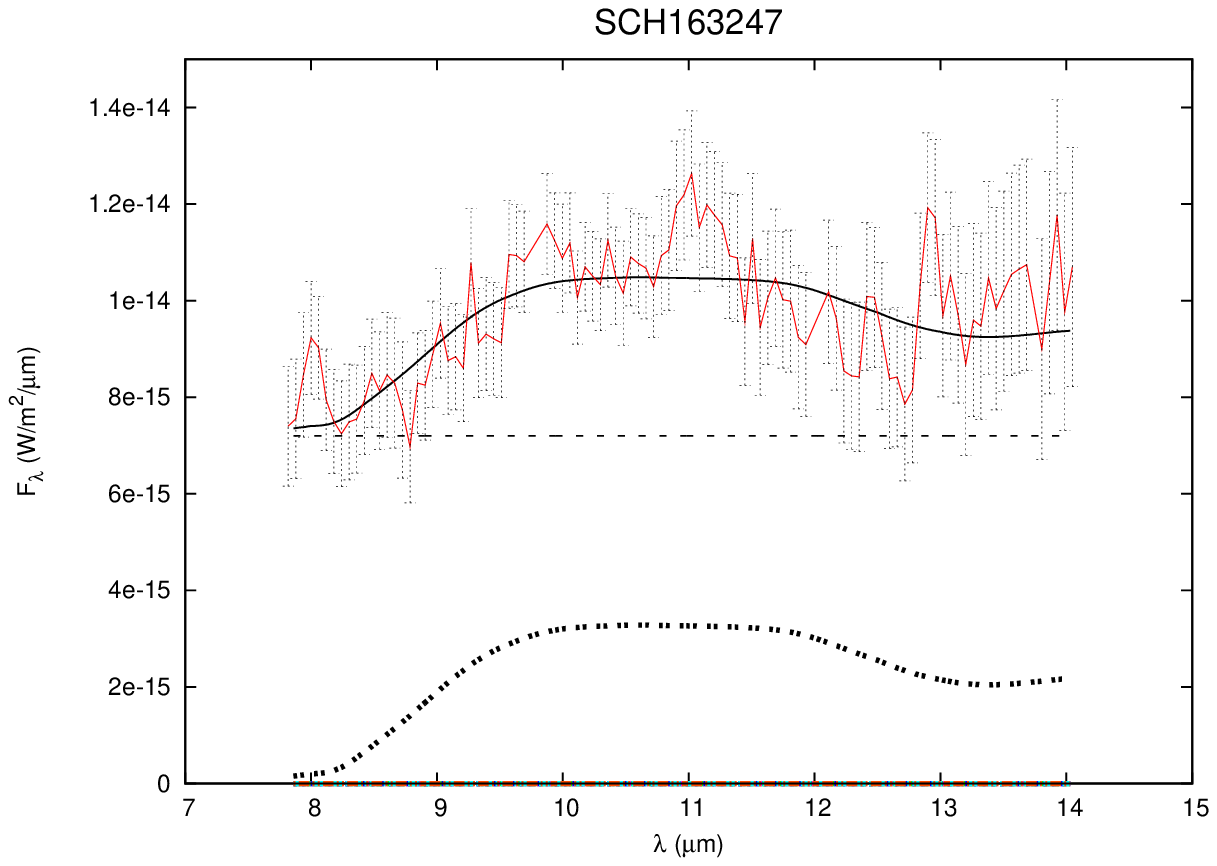}} \\
    \end{tabular}
    \caption{Model-fit to the 10$\micron$ silicate features. Left panel shows the fits obtained using all five dust species, middle panel shows the fits obtained without using any crystalline silicates, right panel shows the fits obtained using only large amorphous olivine and pyroxene grains. Colors represent the following: red--observed spectrum; black--model fit, grey--small amorphous olivine; green--small amorphous pyroxene; cyan--silica; blue--forsterite; pink--enstatite; black dashed--large amorphous olivine; orange--large amorphous pyroxene. Thin dashed line represents the continuum.}
    \label{silicate2}
  \end{center}
 \end{figure}

Figure~\ref{silicate2} shows the compositional fits obtained for SCH162630, usd161939 and SCH163247. Also shown are the additional fits for models (a) and (b). The spectrum for SCH162630 is at a higher S/N of $\sim$25, and the best-fit indicates negligible grain growth in the disk but a high crystalline mass fraction ($\sim$46\%; Table 2). The fits obtained for models (a) and (b) are both of a poor quality for SCH162630 (Fig.~\ref{silicate2}), with the difference in the reduced-$\chi^{2}$ value between the best-fit and these additional fits of $>$0.5. This object has a more flat-peaked spectrum compared to usd161939, a shape which is indicative of a higher degree of dust processing in the disk (e.g., Sargent et al. 2009; R09). The spectra for usd161939 and SCH163247 are noisy (S/N $\sim$ 8). While the fits for model (b) using large amorphous grains are poor for both objects, the fit for model (a) results in a difference in the reduced-$\chi^{2}$ value of only 0.04, indicating crystalline silicates to be unimportant in these disks. However, SCH163247 shows a slight peak near 11.3 $\micron$ which is indicative of crystalline forsterite, and some fraction of this dust species is required in the model to better fit the enhanced emission observed at this wavelength. The mass fractions and the reduced-$\chi^{2}$ values for the best-fits are listed in Table 2. 

\begin{figure}
 \begin{center}
    \begin{tabular}{ccc}      
    \resizebox{80mm}{!}{\includegraphics[angle=0]{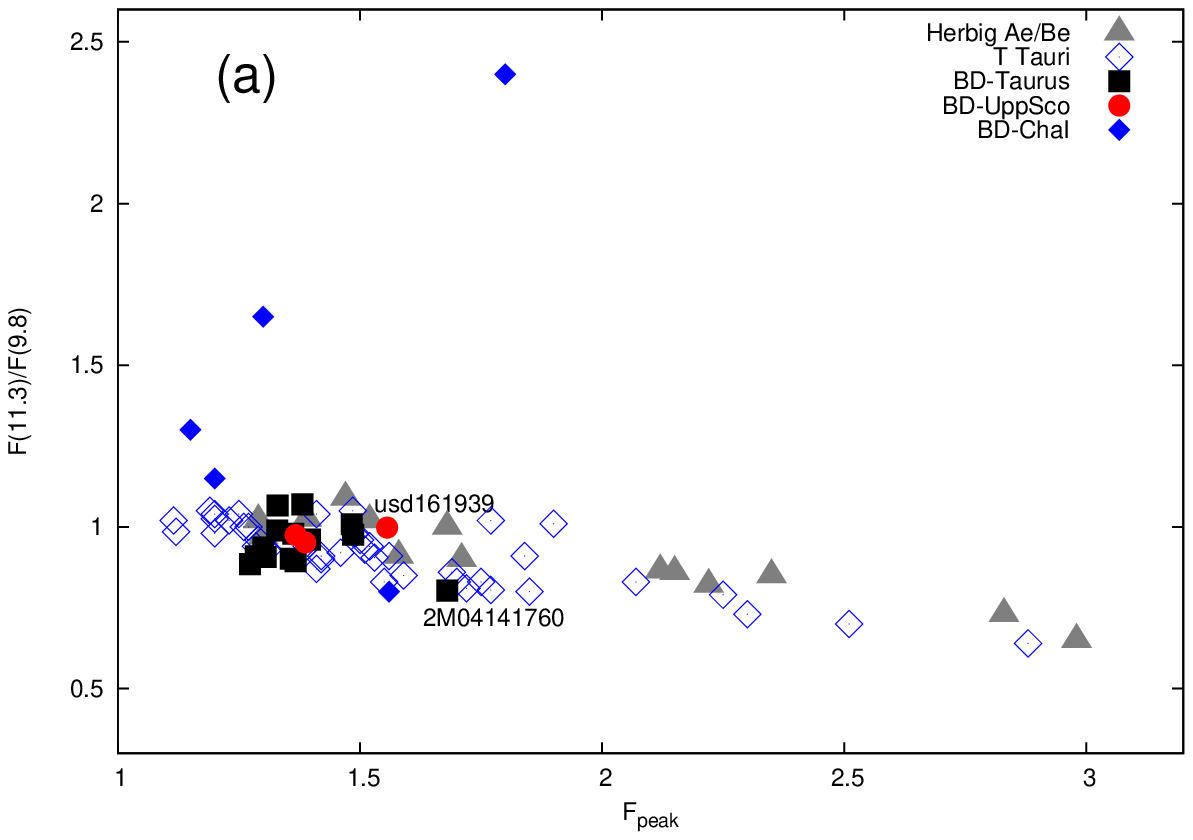}} &  
      \resizebox{80mm}{!}{\includegraphics[angle=0]{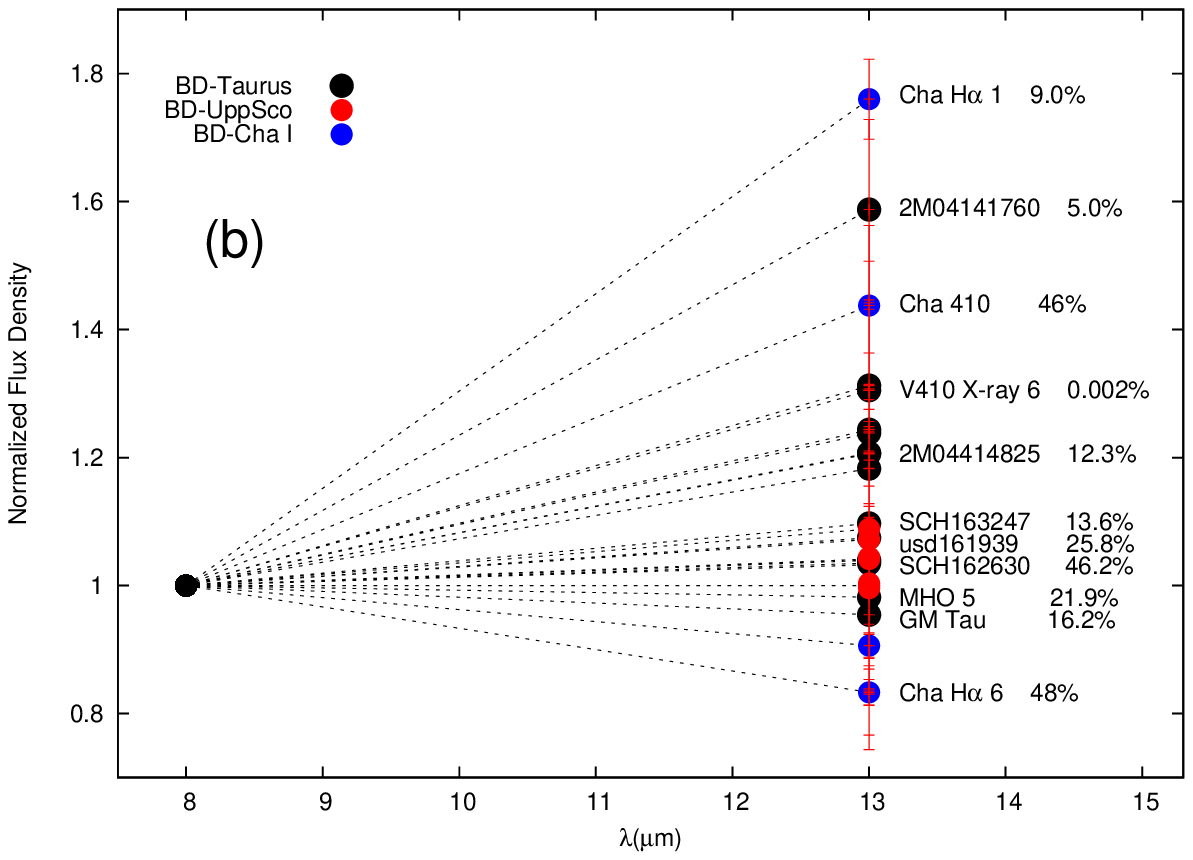}}  \\  
      \resizebox{80mm}{!}{\includegraphics[angle=0]{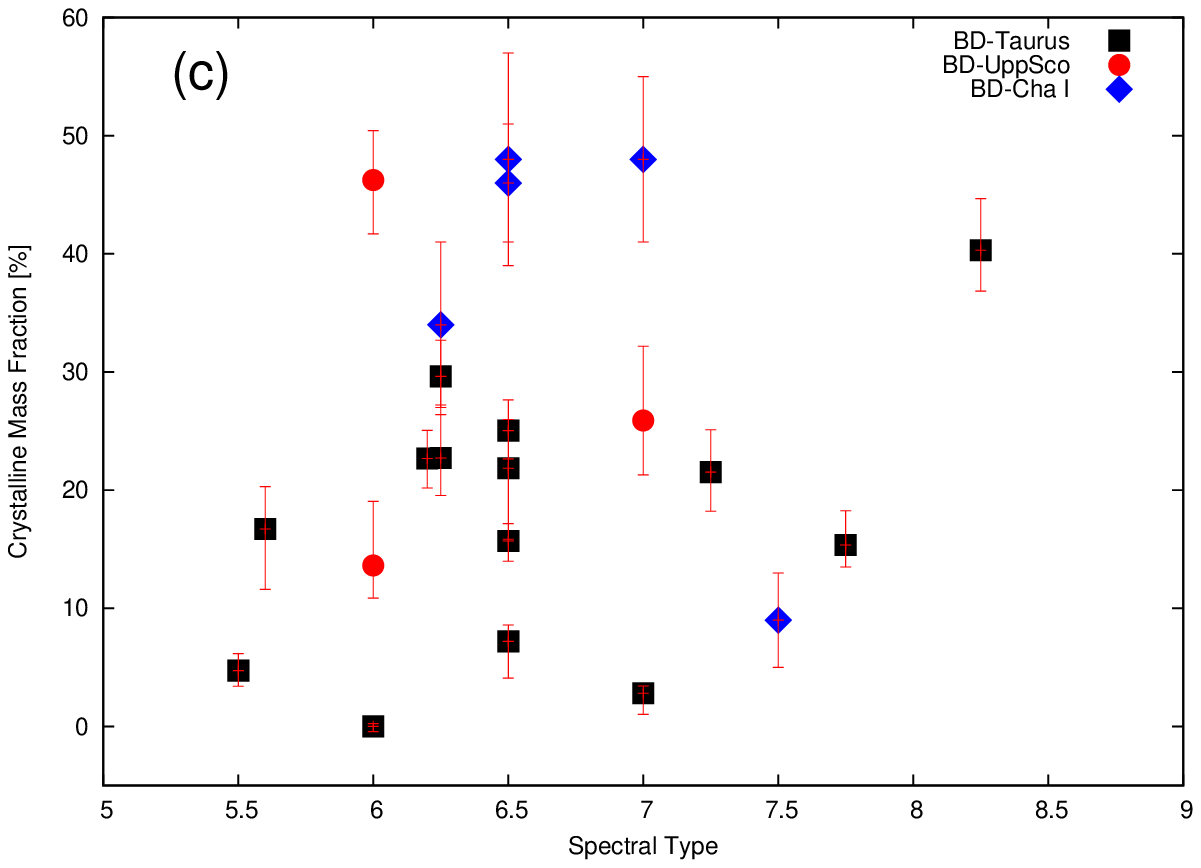}} &
      \resizebox{80mm}{!}{\includegraphics[angle=0]{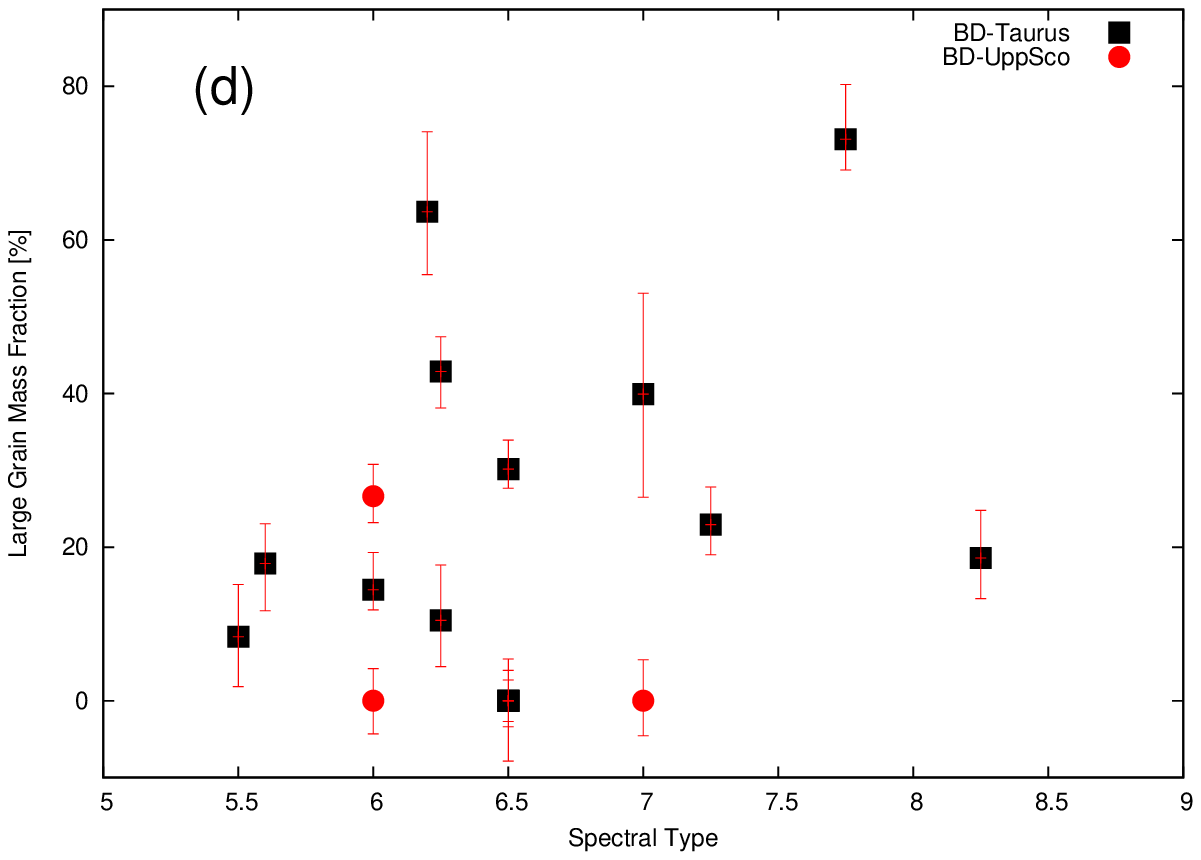}} \\
    \end{tabular}
    \caption{(a) The shape vs. strength of the silicate emission features. Taurus brown dwarfs are indicated by filled black squares, UppSco ones by filled red circles. Included for comparison are T Tauri stars from Pascucci et al. (2009) and Herbig Ae/Be stars from van Boekel et al. (2005). (b): Disk geometry for the Taurus, UppSco and Cha I brown dwarf disks. The values next to the target names are the crystalline mass fractions. (c)-(d): Crystalline and large-grain mass fractions vs. SpT. The values of 5-9 indicate SpTs of M5-M9. }
    \label{plots}
  \end{center}
 \end{figure}

Figure~\ref{plots}a compares the `shape' and `strength' in the 10$\micron$ silicate emission feature for the UppSco, Taurus (R09) and Cha I (Apai et al. 2005) brown dwarfs with higher mass T Tauri stars (Pascucci et al. 2009) and Herbig Ae/Be stars (van Boekel et al. 2005). The feature strength can be estimated from the peak normalized flux, $F_{peak}$, above the continuum, and the shape can be estimated by the ratio of the normalized fluxes at 11.3 and 9.8$\micron$ (e.g., Bouwman et al. 2001). The shapes and strengths for the UppSco brown dwarfs are comparable to the Taurus objects, that cluster around $F_{peak} \sim$ 1.3 and $F_{11.3}/F_{9.8} \sim$ 1.0, with the features being weaker in these objects compared to most other T Tauri and Herbig Ae/Be stars. The brown dwarf usd161939 shows a more narrow, peaked feature than the rest, which is consistent with a high 74\% small amorphous grain mass fraction found for this object. We do not observe in UppSco or Taurus any of the extreme cases as in Cha I that show an increase in the feature strength with increasing crystallinity. This `reversal' in the observed shape vs. strength trend for the Cha I brown dwarfs was explained by Apai et al. (2005) due to the presence of highly processed dust grains in the disk. Considering that none of the Taurus objects show such a behaviour, and the UppSco objects mostly have flat features,  the reversal seems to be occurring at some intermediate age of $\sim$2-3 Myr. Fig.~\ref{plots}b compares the disk structures for the three UppSco brown dwarfs with the ones in the Taurus and Cha I regions. Here, we have plotted the flux densities at 8 and 13$\micron$, normalized to the flux at 8$\micron$. The normalized flux density at 13$\micron$ will be higher for a flared disk, and will decrease as the disk geometry gets flatter. The processing of dust into crystalline silicates as well as growth to larger sizes both effect the vertical structure of the disk. While a few sources in Taurus and Cha I show flared geometries, all three UppSco brown dwarfs show flattened disk structures, comparable to the Taurus objects MHO 5 and GM Tau that are dominated by processed dust ($\sim$20\% crystalline mass fractions; R09). The brown dwarf SCH162630 with a 46\% crystalline mass fraction shows a slightly more flattened disk structure than the other two objects, though any dependence of the crystallinity levels on the extent of dust settling in the disk for the UppSco brown dwarfs is difficult to probe for such a small sample. R09 had noted a lack of any strong correlation between the two processes for the Taurus brown dwarfs. 

Figs.~\ref{plots}c-d compare the crystalline and large-grain mass fractions among the UppSco, Taurus and Cha I brown dwarfs. In the small SpT range probed here, we do not find any particular trends for the crystalline or the large-grain mass fractions. The crystalline mass fractions for the three UppSco sources range between $\sim$14\% and 46\%, with a median value of 26\%. For 15 brown dwarfs in Taurus that show emission in the 10 $\micron$ silicate feature, R09 had reported a 20\% median crystalline mass fraction, comparable to the UppSco sources. However the very small number statistics of 3 for the UppSco objects, and the uncertainties in the mass fractions for usd161939 and SCH163247 due to noisy spectra make any such comparison difficult. All three UppSco sources, on the other hand, have small amorphous grain mass fractions of $>$50\% (Table 2), and smaller large-grain mass fractions compared to most of the Taurus brown dwarfs (Fig.~\ref{plots}d). This may be another indication of the $\sim$5 Myr UppSco objects to be more evolved systems, since increasing grain growth and sedimentation could result in an increase in the fraction of small amorphous grains in the disk surface layers, as the larger grains settle towards the disk midplane (e.g., Dullemond \& Dominik 2008; Honda et al. 2006). 

We can consider three main categories for the silicate emission features: (a) objects with `flat' spectra around 10 $\micron$ with the emission levels consistent with the underlying continuum, (b) objects with `weak' spectra such as usco112 that may be transitional objects, and (c) objects with emission in the silicate feature. Among the 20 Taurus objects studied by R09, there is a $\sim$25\% fraction of flat features, and a nearly equal $\sim$20\% fraction of weak spectra (e.g., CFHT-BD-Tau 20). The detection rate for emission is then about 55\% in Taurus. Combining this work and the S07 survey, we have a total of 16 bona fide disks in UppSco. A 62\% fraction of these show flat spectra (10/16; S07), while about 19\% show weak spectra (Fig.~\ref{silicate}), and a similar fraction is found for the emission spectra. Thus the fraction of flat features has increased by a factor of $\sim$2 over a $\sim$1-5 Myr timescale, while the detection rate in emission has decreased by a similar factor. The fraction of weak features that may be in transition from emission to flat spectra seems comparable in this age range. In the TWA (age of $\sim$10Myr; Webb et al. 1999), the two confirmed disk sources, 2MASSW J1207334-393254 (2M1207) and SSSPM J1102-3431 (SSPM1102), show flat silicate spectra, though slight absorption due to a more edge-on disk is observed for 2M1207 (Riaz \& Gizis 2008). The overall properties of the silicate emission features are thus consistent with a picture of increasing dust processing with age, since such flat spectra are an indication that significant grain growth and dust settling has occurred, and the disks are in a more advanced stage of evolution (e.g., S07; R09).

\subsection{Disk Evolution in Brown Dwarfs}
\label{evolution}

From this archival survey of 28 UppSco brown dwarfs, we estimate a disk fraction of 10.7\%$^{+8.7\%}_{-3.3\%}$ (3/28), where the uncertainties are the 1$\sigma$ Gaussian limits. If we consider the one candidate transition disk, then the fraction increases to 14.3\%$^{+8.7\%}_{-4.3\%}$ (4/28). This fraction is lower by a factor of $\sim$3 compared to the 37\%$\pm$9\% (13/35) reported by S07. Combining the two surveys, we have a fraction of 27\%$^{+6.0\%}_{-4.8\%}$ (17/63) for brown dwarf disks in UppSco. In the TWA, out of the four confirmed brown dwarf members, two have been confirmed as disk-bearing objects (2M1207 and SSPM1102; Sterzik et al. 2004; Riaz \& Gizis 2008), resulting in a 50\%$\pm$20\% disk fraction. The TWA disk fraction thus could be somewhere between 30\% and 70\%. This suggests a constant brown dwarf disk fraction over a $\sim$5-10 Myr timescale, but could also suggest an increase in the fraction by a factor of $\sim$2. At younger ages, Luhman et al. (2006) have reported a 40\%$\pm$13\% brown dwarf disk fraction in Taurus ($\sim$1Myr), while the fractions are 50\% $\pm$ 17\% and 42\% $\pm$ 13\% in the $\sim$2-3Myr Cha I and IC 348 regions, respectively (Luhman et al. 2005). Among earlier type stars, Luhman et al. (2005) have reported disk fractions of 45\% $\pm$ 7\% and 33\% $\pm$ 4\%, respectively, in the Cha I and IC 348 regions for the M0-M6 members (0.7 $M_{\odot}$ $\geq$ {\it M} $\geq$ 0.1 $M_{\odot}$). Carpenter et al. (2006) have reported a 19\% $\pm$ 5\% fraction for the K0-M5 stars in UppSco, while Low et al. (2005) have found a disk fraction of 24\%$^{+9\%}_{-6.4\%}$ (6/25) for the stellar members in the TWA. If we consider only the M0-M6 stars, then the TWA disk fraction drops down to 16.6\%$^{+11.7}_{-5.4\%}$ (3/18). A simple comparison then suggests that the brown dwarf disk fraction has not decreased by an age of $\sim$10 Myr, while it shows a drop by a factor of $\sim$2 among the higher mass stars. Brown dwarf disks thus may survive until $\sim$10 Myr, as earlier noted by e.g., S07, Bouy et al. (2007) and Riaz \& Gizis (2008). The current statistics in the TWA are too limited to put any limits on the disk evolutionary timescale for brown dwarfs. Also, all such conclusions are affected by the uncertainty on the age estimate for any given cluster. 

There is one source SCH160930 that may be a candidate transition disk, resulting in a fraction of 3.6\%$^{+7}_{-1.1}$ of such systems among the UppSco brown dwarfs. This is consistent within the uncertainties with the 8\% upper limit estimated by S07 for brown dwarf transition disks in UppSco. SCH160930 shows photospheric emission shortward of 24$\micron$, indicating an inner disk hole as large as $\sim$1 AU. Such large inner holes have been reported earlier for brown dwarfs in the IC 348 cluster (Muzerolle et al. 2006). The absence of excess emission at shorter wavelengths could be either due to grain removal processes such as photoevaporation, or substantial grain growth and dust settling in the inner disk regions (e.g., Muzerolle et al. 2006; Cieza et al. 2007).

\begin{figure}
 \begin{center}
    \begin{tabular}{ccc}      
    \resizebox{170mm}{!}{\includegraphics[angle=0]{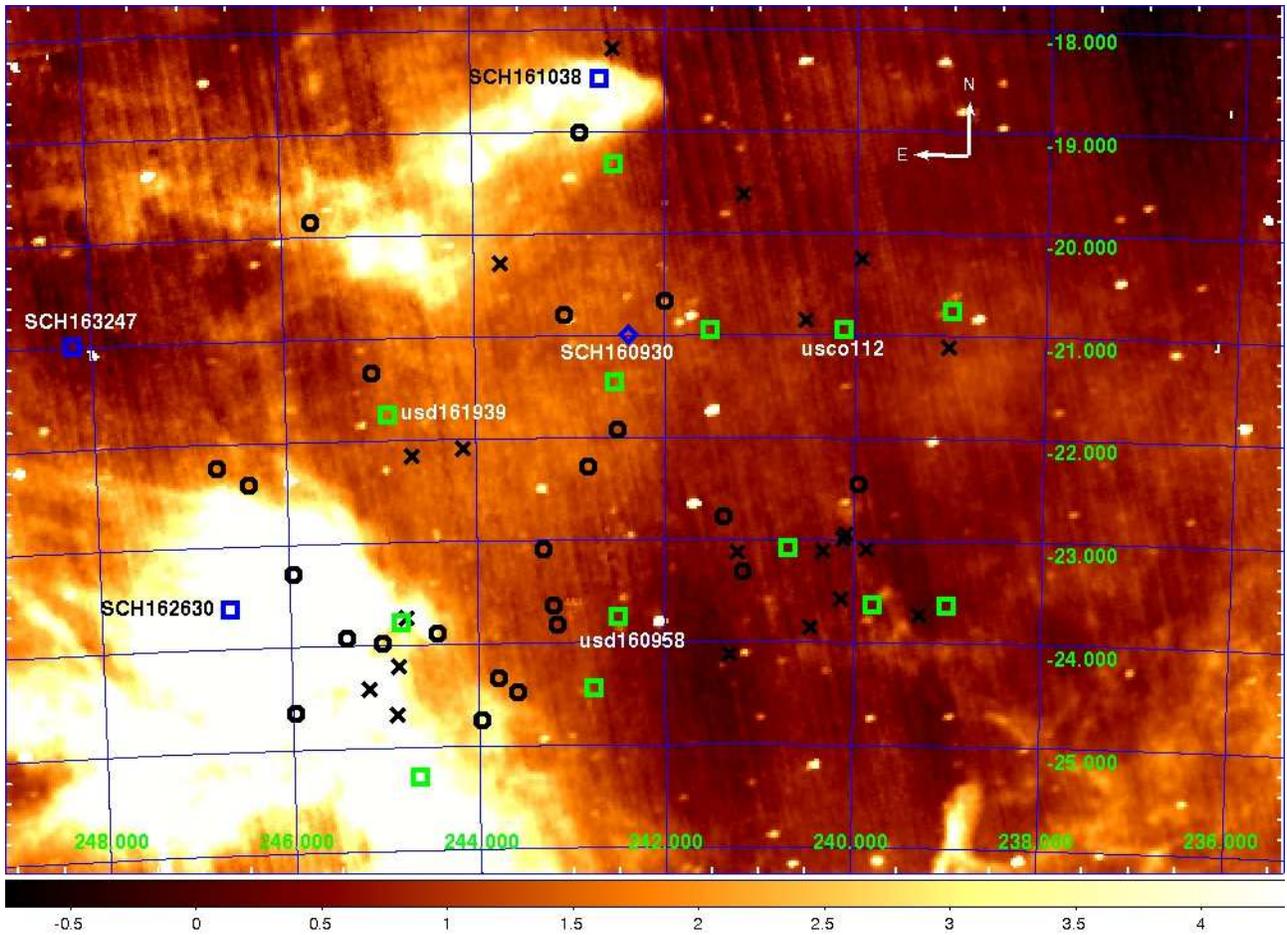}} \\
        \end{tabular}
    \caption{Spatial distribution for the disk-bearing objects from this work and the S07 survey. The image is a false-color 13$\degr$$\times$8$\degr$ {\it IRAS} 12 $\micron$ observation. The color intensity scale in units of MJy sr$^{-1}$ is shown at the bottom, and the coordinate grid in degrees (J2000) is overlaid. Symbols represent: black circles--S06 brown dwarfs without disks; black crosses--S07 objects without disks; blue squares and blue diamond represent the three disk-bearing objects and the one candidate transition disk, respectively, found in this work; green squares represent the disk-bearing objects from the S07 survey.   }
    \label{image}
  \end{center}
 \end{figure}

The disk fraction obtained here is a factor of $\sim$3 smaller than that reported by S07 for brown dwarfs in UppSco. In Fig.~\ref{image}, we have examined the spatial distribution for the disk-bearing sources in the S07 sample and in this work. While 80\% of the S06 sources have $A_{v} <$ 1, the source SCH162630 lies in a more extincted region towards the south-east, and S06 have measured an $A_{v}$ of 3.02 for this object. SCH161038 that lies in the northern region also shows some extinction of $A_{v}$=1.78 (S06). The third disk SCH163247 however lies in a clear region with an $A_{v}$ of 0.06. While these three sources lie to the east of most disk-bearing objects from the S07 sample, we do not see any particular differences in the spatial distribution of the disk sources in the two samples, that might indicate an age effect. 

We consider the brown dwarf densities in the Taurus, UppSco and TWA regions, that might affect the disk fractions noted above. According to Briceno et al. (2002) and Luhman et al. (2003), the number of brown dwarfs in Taurus with SpT between M6 and M9 and $A_{v} <$ 4 mag is 10 over 8.4 square degrees, implying a density of $\sim$ 1 object per square degree. For UppSco, S06 confirmed 43 members over 200 square degrees from a sample of 20\% of all photometric candidates. Out of those 43 sources, 30 are brown dwarfs (SpT$\ga$M6). Thus the total survey should contain about 5x30=150 objects in 200 square degrees that have masses between 75 and 20 Jupiter masses, implying a density of 0.75 per square degree. From the work of Lodieu et al. (2007), there are 23 brown dwarfs with estimated masses between 75 and 20 Jupiter masses over 6.5 square degrees, implying a density of 3.5 brown dwarfs per squre degree. Their survey however is targeted more towards the core of UppSco, which could explain the higher densities. The difference with Taurus in terms of density is thus not very large and of the same order. This is consistent with the comparable disk fractions observed in the two regions ($\sim$30\% in UppSco, $\sim$40\% in Taurus). In comparison, the TWA is very loose and contains a small number of members (e.g., Webb et al. 1999). Following Jayawardhana et al. (1999), the number of young stellar objects belonging to the TWA in the solar neighbourhood would be of the order of 0.06 objects per square degree or less. Considering a typical value of $\sim$5 for the ratio of stars to sub-stellar objects in young nearby clusters (Andersen et al. 2008), we can estimate a brown dwarf density in the TWA that is at least $\sim$100 times less than that in Taurus or UppSco. The TWA however is a factor of $\sim$3 closer than Taurus or UppSco, and therefore even with the same object density per cubic parsec, we would expect a drop by a factor of $\sim$10 in the density per square degree. Although it can be argued that the larger age of the TWA allows more time for the dispersion of the young sources, the much lower densities in this region could explain the longer disk dissipation timescales, and disk lifetimes may be prolonged in a less dense environment.

\section{Summary}
We report a disk fraction of 10.7\%$^{+8.7\%}_{-3.3\%}$ from a {\it Spitzer} archival survey of 28 UppSco brown dwarfs. One object shows a small excess at 24 $\micron$ but none at shorter wavelengths, and may be a candidate transition disk. The fraction of flat silicate spectra has increased by a factor of $\sim$2 over a $\sim$1-5 Myr timescale. While a few Taurus sources show flaring in the disk, all disk-bearing brown dwarfs in UppSco and the TWA show flat disk structures, consistent with a picture of increasing dust processing and dust settling with age. We do not find any differences in the spatial distributions for the disk-bearing objects found in this work, and from the S07 survey, that might indicate an age spread in the UppSco. The much lower brown dwarf densities in the TWA might be an explanation for the longer disk lifetimes observed among the sub-stellar objects in this region. 

\acknowledgements
We thank the anonymous referee for helpful comments and suggestions. Support for this work was provided by CONSTELLATION grant \# YA 2007. NL is funded by the Ram\'on y Cajal fellowship number 08-303-01-02\@. This work is based on observations made with the {\it Spitzer Space Telescope}, which is operated by the Jet Propulsion Laboratory, California Institute of Technology under a contract with NASA.

\clearpage

\begin{deluxetable}{cccccccccccccc}
\tabletypesize{\scriptsize}
\tablecaption{IRAC and MIPS magnitudes}
\tablewidth{0pt}
\tablehead{
\colhead{Name}  & \colhead{$\alpha$ (J2000)} & \colhead{$\delta$ (J2000)} & \colhead{[3.6]} & \colhead{[4.5]} & \colhead{[5.8]} & \colhead{[8]} & \colhead{[24]} & \colhead{[70]}
} 
\startdata

SCH155948 & 15 59 48.02 &  -22 27 16.5 &  12.77 (0.04) & 12.62 (0.04) & 12.55 (0.05) & 12.51 (0.07) & -- & -- \\
SCH160443 & 16 04 43.03  & -23 18 26.2 &  12.32 (0.03) & 12.25 (0.04) & 12.30 (0.05) & 12.19 (0.07) & -- & -- \\
SCH160530 & 16 05 30.77 &  -22 46 20.2 &  12.40 (0.03) & 12.40 (0.04) & 12.32 (0.05) & 12.34 (0.06) & -- & -- \\
SCH160758 & 16 07 58.50 &  -20 39 48.9 &  12.25 (0.03) & 12.09 (0.03) & 11.47 (0.06) & 11.33 (0.07) & -- & -- \\
SCH160930 & 16 09 30.18 &  -20 59 54.1 &  12.63 (0.04) & 12.55 (0.04) & 12.54 (0.05) & 12.32 (0.06) & 8.88 (0.23) & -- \\
SCH160959 & 16 09 59.91 &  -21 55 42.9 &  12.95 (0.04) & 12.87 (0.05) & 12.77 (0.06) & 12.72 (0.08) & -- & -- \\
SCH161038 & 16 10 38.76 &  -18 29 23.5 &  12.07 (0.03) & 11.78 (0.03) & 11.42 (0.03) & 10.83 (0.03) & 7.19 (0.15) & -- \\
SCH161117 & 16 11 17.11 &  -22 17 17.5 &  12.92 (0.04) & 12.87 (0.05) & 12.74 (0.06) & 12.86 (0.09) & -- & -- \\
SCH161129 & 16 11 29.59 &  -19 00 29.2 &  12.04 (0.03) & 11.90 (0.03) & 12.02 (0.06) & 11.79 (0.06) & -- & -- \\
SCH161211 & 16 12 11.88 &  -20 47 27.0 &  12.23 (0.03) & 12.12 (0.03) & 12.06 (0.04) & 12.05 (0.06) & -- & -- \\
SCH161237 & 16 12 37.58 &  -23 49 23.4 &  12.58 (0.03) & 12.54 (0.04) & 12.44 (0.05) & 12.50 (0.07) & -- & -- \\
SCH161246 & 16 12 46.92 &  -23 38 40.9 &  12.28 (0.03) & 12.22 (0.03) & 12.16 (0.04) & 12.10 (0.06) & -- & -- \\
SCH161312 & 16 13 12.12 &  -23 05 03.3 &  12.60 (0.04) & 12.55 (0.04) & 12.44 (0.05) & 12.44 (0.07) & -- & -- \\
SCH161419 & 16 14 19.74 &  -24 28 40.5 &  12.36 (0.03) & 12.30 (0.04) & 12.26 (0.05) & 12.13 (0.06) & -- & -- \\
SCH161511 & 16 15 11.15 &  -24 20 15.6 &  12.85 (0.04) & 12.82 (0.05) & 12.69 (0.05) & 12.69 (0.08) & -- & -- \\
SCH161555 & 16 15 55.08 &  -24 44 36.8 &  11.92 (0.03) & 11.88 (0.03) & 12.00 (0.06) & 11.75 (0.07) & -- & -- \\
SCH161745 & 16 17 45.40 &  -23 53 36.2 &  12.56 (0.03) & 12.50 (0.04) & 12.45 (0.05) & 12.50 (0.08) & -- & -- \\
SCH162007 & 16 20 07.56 &  -23 59 15.2 &  11.67 (0.02) & 11.59 (0.03) & 11.58 (0.05) & 11.26 (0.05) & -- & -- \\
SCH162021 & 16 20 21.27 &  -21 20 29.2 &  12.01 (0.03) & 11.88 (0.03) & 11.74 (0.07) & 11.94 (0.09) & -- & -- \\
SCH162243 & 16 22 43.84 &  -19 51 05.8 &  10.56 (0.01) & 10.48 (0.02) & 10.43 (0.03) & 10.38 (0.03) & -- & -- \\
SCH162528 & 16 25 28.62 &  -16 58 50.5 &  12.24 (0.03) & 12.16 (0.03) & 12.09 (0.04) & 12.02 (0.06) & -- & -- \\
SCH162536 & 16 25 36.71 &  -22 24 28.9 &  11.99 (0.03) & 11.94 (0.03) & 11.90 (0.05) & 11.75 (0.06) & -- & -- \\
SCH162630 & 16 26 30.26 &  -23 36 55.5 &  11.30 (0.02) & 10.95 (0.02) & 10.64 (0.03) &  9.95 (0.02) & 6.43 (0.12) & -- \\
SCH162656 & 16 26 56.19 &  -22 13 52.2 &  12.08 (0.03) & 11.99 (0.03) & 11.80 (0.06) & 12.09 (0.09) & -- & -- \\
SCH163247 & 16 32 47.26 &  -20 59 37.7 &  11.94 (0.03) & 11.70 (0.03) & 11.43 (0.03) & 10.92 (0.03) & 6.62 (0.07) & -- \\
2M1139 & 11 39 51.14 &  -31 59 21.5 &  10.91 (0.06) & 10.81 (0.07) & 10.72 (0.11) & 10.53 (0.12) & 9.53 (0.22) & 5.70\tablenotemark{a} (0.4) \\

\enddata
\tablenotetext{a}{The 2$\sigma$ upper limit.}
\end{deluxetable}

\begin{deluxetable}{cccccccccccccc}
\tabletypesize{\scriptsize}
\tablecaption{Model Fit Parameters}
\tablewidth{0pt}
\tablehead{
\colhead{Name} & \colhead{$F_{peak}$} & \colhead{$F_{11.3}/F_{9.8}$} & \colhead{Amorphous\tablenotemark{a}} & \colhead{Large\tablenotemark{b}} & \colhead{Crystalline\tablenotemark{c}} &\colhead{$\chi^{2}$ \tablenotemark{d}}  \\
&&& \colhead{Silicates} & \colhead{Silicates} & \colhead{Silicates} \\
} 
\startdata

SCH161038 & -- & -- & -- & -- & -- &--  \\
usd161939 & 1.56 & 1.0 & 74.1\%$^{+15.6\%}_{-18.2\%}$ & 0.005\%$^{+5.3\%}_{-4.5\%}$ & 25.9\%$^{+6.3\%}_{-4.7\%}$ &1.6 \\
SCH162630 & 1.38 & 0.68 & 48.5\%$^{+4.8\%}_{-5.7\%}$ & 0.004\%$\pm$4\% & 46.2\%$^{+4.2\%}_{-5.2\%}$  & 1.1 \\
SCH163247 & 1.37 & 0.7 & 59.7$^{+8.2\%}_{-13.6\%}$ & 26.6\%$^{+3.0\%}_{-3.2\%}$ & 13.6\%$^{+7.2\%}_{-3.3\%}$ & 1.4 \\

\enddata

\tablenotetext{a}{Percentage of small amorphous olivine and pyroxene silicates compared to all other silicates.}
\tablenotetext{b}{Percentage of large amorphous olivine and pyroxene silicates compared to all other silicates.}
\tablenotetext{c}{Percentage of crystalline silicates (enstatite and forsterite) compared to all other silicates.}
\tablenotetext{d}{The reduced-$\chi^{2}$ value of the best-fit.}

\end{deluxetable}

\end{document}